\documentclass[aps,twocolumn,showpacs,floatfix]{revtex4}

\usepackage[usenames]{color}

\usepackage{amssymb,epsfig}

\usepackage{psfrag}

\renewcommand{\theequation}{\arabic{section}.\arabic{equation}}

\newcommand{\beq}[1]{
\begin{equation}\label{#1}}
\newcommand{\eeq}{\end{equation}}
\newcommand{\bea}[1]{
\begin{eqnarray}\label{#1}}
\newcommand{\eea}{\end{eqnarray}}

\newcommand{\bra}[1]{\left\langle #1 \right|}
\newcommand{\ket}[1]{\left| #1 \right\rangle}

\newcommand{\ep}{\varepsilon}
\newcommand{\ga}{\gamma}

\newcommand{\dd}{{\rm d}}

\newcommand{\nn}{\nonumber}

\psfrag{Q2}{{\scriptsize $Q^2$,~GeV$^2$}}

\psfrag{G1pGD}{{\scriptsize $Q^2G^{\pi^0 p}_1/(m_N^2G_D)$}}

\psfrag{G2pGD}{{\scriptsize $G^{\pi^0 p}_2/G_D$}}

\psfrag{G1nGD}{{\scriptsize $Q^2G^{\pi^+ n}_1/(m_N^2G_D)$}}

\psfrag{G2nGD}{{\scriptsize $G^{\pi^+ n}_2/G_D$}}

\psfrag{G1npGD}{{\scriptsize $Q^2G^{\pi^- p}_1/(m_N^2G_D)$}}

\psfrag{G2npGD}{{\scriptsize $G^{\pi^- p}_2/G_D$}}

\psfrag{G1nnGD}{{\scriptsize $Q^2G^{\pi^0 n}_1/(m_N^2G_D)$}}

\psfrag{G2nnGD}{{\scriptsize $G^{\pi^0 n}_2/G_D$}}

\psfrag{EpGd}{{\scriptsize $E^{\pi^0 p}_{0+}/G_D$}}

\psfrag{LpGd}{{\scriptsize $L^{\pi^0 p}_{0+}/G_D$}}

\psfrag{EnGd}{{\scriptsize $E^{\pi^+ n}_{0+}/G_D$}}

\psfrag{LnGd}{{\scriptsize $L^{\pi^+ n}_{0+}/G_D$}}

\psfrag{pin}{{\scriptsize $\pi^+n$}}

\psfrag{pip}{{\scriptsize $\pi^0p$}}

\psfrag{pin}{{\scriptsize $\pi^+n$}}

\psfrag{pip}{{\scriptsize $\pi^0p$}}

\psfrag{G1GA}{{\scriptsize $\sqrt{2}Q^2G_1^{\pi N}/(G_A m_N^2)$}}

\psfrag{G2GA}{{\scriptsize $\sqrt{2}|G_2^{\pi N}|/G_A$}}

\psfrag{F2}{{\scriptsize $10^3\times F_2^p(W,Q^2)$}}

\psfrag{W2}{{\scriptsize $W^2$,~GeV$^2$}}

\psfrag{pi0piplus}{{\scriptsize $F_2^{\gamma^*p\to \pi^0 p}/F_2^{\gamma^*p\to X}$}}

\psfrag{dsigma}{{\scriptsize $Q^6 \sigma_{\gamma^*p\to \pi^0 p}$}}

\psfrag{ds}{{\footnotesize $d\sigma_{\gamma^*p\to \pi^0 p}/d\Omega_\pi,~\mu\mbox{b/ster}$}}

\psfrag{cost}{{\footnotesize $\cos\theta$}}

\begin{document}


\preprint@{
\hspace{2.2mm}IPPP/07/65\\
\phantom{abc}DCPT/07/130\\
}
\vspace{4mm}

\title{Threshold Pion Electroproduction at Large Momentum Transfers}

\author{V. M. Braun$^1$, D. Yu. Ivanov$^2$
and A. Peters$^1$\\}

\affiliation{
{}$^1$ Institut f\"ur Theoretische Physik, Universit\"at
          Regensburg, D-93040 Regensburg, Germany \\
{}$^2$ Sobolev Institute of Mathematics, 630090 Novosibirsk, Russia}

\date{\today}

\begin{abstract}
We consider pion electroproduction close to threshold for $Q^2$ in the region $1-10$~GeV$^2$
on a nucleon target.
The momentum transfer dependence of the
S-wave multipoles at threshold, $E_{0+}$ and $L_{0+}$, is calculated
in the chiral limit using light-cone sum rules.
Predictions for the cross sections in the threshold region are given
taking into account P-wave contributions that, as we argue, are
model independent to a large extent. The results are compared with
the SLAC E136 data on the structure function $F_2(W,Q^2)$
in the threshold region.
\end{abstract}

\pacs{12.38.-t, 14.20.Dh; 13.40.Gp}




\maketitle


\section{Introduction}
\setcounter{equation}{0}

Threshold pion photo- and electroproduction $\gamma N \to \pi N$, $\gamma^\ast N \to \pi N$ is a 
very old subject that has been receiving continuous attention from both experimental and theoretical
side for many years. From the theory point of view, the interest is because in the approximation of the
vanishing pion mass chiral symmetry supplemented by current algebra allow one
to make exact predictions for the threshold cross sections, known as
low-energy theorems (LET) \cite{KR,Nambu:1997wa,Nambu:1997wb}. As a prominent example, the LET establishes a connection
between charged pion electroproduction and the axial form factor of the nucleon.
In the real world the pion has a mass, $m_\pi/m_N\sim 1/7$, and the study of finite pion mass corrections to LET
was a topical field in high energy physics in the late sixties and early seventies before the celebrated discovery
of Bjorken scaling in deep--inelastic scattering and the advent of QCD, see, in particular, the
work by Vainshtein and Zakharov \cite{Vainshtein:1972ih} and a monograph by Amaldi, Fubini and Furlan \cite{AFF}
that addresses many of these developments.

Twenty years later, a renewed interest to threshold pion production was trigged by the extensive data that became
available on $\gamma p \to \pi^0 p$ \cite{Mazzucato:1986dz,Beck:1990da} and, most importantly, $\gamma^* p \to \pi^0 p$,
at the photon virtuality $Q^2 \sim 0.04-0.1$~GeV$^2$ \cite{Welch:1992ex}. At the same time, the advent of
chiral perturbation theory (CHPT) has allowed for the systematic expansion of low--energy physical observables
in powers of the pion mass and momentum. In particular classic LET were reconsidered and rederived in this new
framework, putting them on a rigorous footing, see \cite{Bernard:1995dp} for an excellent review.
The new insight brought by CHPT calculations is that certain loop
diagrams produce non-analytic contributions to scattering amplitudes that are lost in the naive expansion
in the pion mass, e.g. in \cite{Vainshtein:1972ih,Scherer:1991cy}. By the same reason, the expansion at small
photon virtualities $Q^2$ has to be done with care as the limits $m_\pi\to0$ and $Q^2\to 0$ do not commute,
in general \cite{Bernard:1992ys}. The LET predictions including CHPT corrections
seem to be in good agreement with experimental data on pion photoproduction \cite{Drechsel:1992pn}.
Experimental results on the S-wave electroproduction cross section for  $Q^2 \sim 0.1$~GeV$^2$
are consistent with CHPT calculations as well, \cite{Bernard:1992rf,Bernard:1995dp}, and
cannot be explained without taking into account chiral loops.

The rapid development of experimental techniques is making possible to study threshold pion production
in high-energy experiments and in particular electroproduction with photon virtuality $Q^2$ in a few GeV$^2$
range. Such experiments would be a major step forward and require very fine energy resolution
in order to come close to the production threshold to suppress the P-wave contribution of the $M_{1+}$ multipole.
Various polarisation measurements can be especially helpful in this respect. We believe that
such studies are feasible on the existing and planned accelerator facilities,
especially at JLAB, and the task of this paper is to provide one with the necessary theoretical guidance.

In the traditional derivation of LET using PCAC and current algebra $Q^2$ is not assumed to be small but
the expansion in powers of the pion mass involves two parameters:
$m_\pi/m_N$ and $m_\pi Q^2/m_N^3$ \cite{Vainshtein:1972ih,Scherer:1991cy}.
The appearance of the second parameter in this particular combination reflects the fact that, for finite pion
masses and large momentum transfers, the emitted pion cannot be 'soft' with respect to the
initial and final state nucleons simultaneously. For the threshold kinematics, this affects in particular the
contribution of pion emission from the initial state \cite{PPS01}
and in fact $m_\pi [Q^2+2m_N^2]/m_N^3$ is nothing but the nucleon virtuality after the pion emission, divided by $m_N^2$.
It follows that the LET are formally valid (modulo CHPT loop corrections \cite{Bernard:1995dp}) for the momentum
transfers as large as $Q^2 \sim m_N^2$ where CHPT is no more applicable, at least in its standard form.
To the best of our knowledge, there has been no dedicated analysis of the threshold production in 
the $Q^2\sim 1$~GeV$^2$ region, however.

{} For $m_\pi Q^2/m_N^3 = {\mathcal O}(1)$ the LET break down: the initial state pion radiation occurs at time scales
of order $1/m_N$ rather than $1/m_\pi$ necessitating to add  contributions of hadronic intermediate states other than
the nucleon. Finally, for very large momentum transfers,
the situation may again become tractable as one can try to separate contributions of
'hard' scales
as coefficient functions in front
of 'soft' contributions involving small momenta and use current algebra (or CHPT) for the latter but not for
the amplitude as a whole.

This approach was pioneered in the present context in  Ref.~\cite{PPS01} where it was suggested
that for asymptotically large $Q^2$ the standard pQCD collinear factorisation technique \cite{Efremov:1979qk,Lepage:1980fj}
becomes applicable and the helicity-conserving $E_{0+}$ multipoles can be calculated (at least for $m_\pi=0$)
in terms of chirally rotated nucleon distribution amplitudes.
In practice one expects that the onset of the pQCD regime is postponed
to very large momentum transfers because the factorisable contribution involves a small factor $(\alpha_s(Q)/2\pi)^2$
and has to win over nonperturbative ``soft'' contributions that are suppressed by an extra power of $Q^2$ but do not
involve small coefficients.

The purpose of this paper is to suggest a realistic QCD-motivated model for the $Q^2$ dependence of
both transverse $E_{0+}$ and longitudinal $L_{0+}$
S-wave multipoles at threshold in the region $Q^2 \sim 1-10$~GeV$^2$ that can be accessed experimentally
at present or in near future.
In Ref.~\cite{Braun:2001tj} we have developed a technique  to calculate baryon form factors
for moderately large $Q^2$ using light-cone  sum rules (LCSR) \cite{Balitsky:1989ry,Chernyak:1990ag}.
This approach is attractive because in LCSR  ``soft'' contributions to the form factors are calculated in
terms of the same nucleon distribution amplitudes (DAs) that
enter the pQCD calculation and there is no double counting. Thus, the LCSR provide one with the most
direct relation of the hadron form factors and distribution amplitudes that is available at present,
with no other nonperturbative parameters.

The same technique can be applied to pion electroproduction.
In Ref.~\cite{Braun:2006td} the relevant generalised form factors were estimated in the LCSR approach for
the range of momentum transfers $Q^2 \sim 5-10$~GeV$^2$.
{}For this work,  we have reanalysed the sum rules derived in \cite{Braun:2006td} taking into account the
semi-disconnected pion-nucleon contributions in the intermediate state. We demonstrate that, with this addition,
the applicability of the sum rules can be extended to the lower $Q^2$ region and the LET  are indeed
reproduced at $Q^2\sim 1$~GeV$^2$ to the required accuracy $\mathcal{O}(m_\pi)$.
The results presented in this work essentially interpolate between the large-$Q^2$ limit considered in
\cite{Braun:2006td} and the standard LET predictions at low momentum transfers.

The presentation is organised as follows.
Section~2 is introductory and contains the necessary kinematics and notations.
In Section~3 we define two generalised form factors that contribute to pion electroproduction at the kinematic threshold,
explain the relation to S-wave multipoles and suggest a model for their $Q^2$ dependence based on LCSR.
The details of the LCSR calculation are presented in the Appendix.
In Section~4 we suggest a simple model for the electroproduction close to threshold, complementing the S-wave
form factor-like contributions by P-wave terms corresponding to pion emission in the final state that can be
expressed in terms of the nucleon electromagnetic form factors. In this framework, detailed predictions are worked
out for the differential cross sections from the proton target and also for the structure functions measured in
the deep-inelastic scattering experiments. The comparison with SLAC E136 results \cite{Bosted:1993cc} is presented.
The final Section~5 is reserved for a summary and conclusions.

\section{Kinematics and Notations}
\setcounter{equation}{0}
{}For definiteness we consider pion electroproduction from a proton target
\bea{piprod}
  e(l)+p(P) &\to& e(l') + \pi^+(k) + n(P')\,,
\nonumber\\
  e(l)+p(P) &\to& e(l') + \pi^0(k) + p(P')\,.
\eea
Basic kinematic variables are
\begin{eqnarray}
&&q=l-l'\, , \quad s=(l+P)^2, \quad W^2=(k+P')^2 \,,
\nonumber\\
&& q^2=-Q^2 \, , \quad P'^2=P^2=m_N^2 \, , \quad k^2=m^2_\pi\,,
\nonumber \\
&& y=\frac{
P\cdot q}{ P\cdot l}=\frac{W^2+Q^2-m_N^2}{ s-m_N^2}\,.
\end{eqnarray}
The identification of the momenta is clear from Eq.~(\ref{piprod});
$m_N$ is the nucleon and $m_\pi$ the pion mass, respectively.
In what follows we neglect the electron mass and the difference of proton and neutron masses.

The differential cross section for electron scattering in laboratory frame is equal to
\beq{labsec}
\frac{d\sigma}{dE' d \Omega'} = \left(\frac{E'}{E}\right)\frac{\beta(W) \,d\Omega_\pi}{64 m_N (2\pi)^5}
\,\frac{4\pi \alpha_{\rm em}}{Q^4} L_{\mu\nu}M^{\mu\nu}.
\eeq
Here
\bea{lephadtensors}
 L_{\mu\nu} &=& (\bar u(l') \gamma_\mu u(l) ) (\bar u(l') \gamma_\mu u(l) )^\ast ,
\nonumber\\
 M^{\mu\nu} &=& 4\pi \alpha_{\rm em} \langle N\pi|j_\mu^{em}|p\rangle  \langle N\pi|j_\nu^{em}|p\rangle^\ast,
\eea
where the sum (average) over the polarisations is implied,
$ d\Omega_\pi = d\phi_\pi d(\cos\theta)$,
$\theta$ and $ \phi_\pi$ being the polar and azimuthal angles of the pion in the
final nucleon-pion c.m. frame, respectively,
the electromagnetic current is defined as
\beq{jmu}
 j^{\mathrm{em}}_\mu (x) = e_u\bar u(x) \gamma_\mu u(x) + e_d\bar d(x) \gamma_\mu d(x)\,
\eeq
and $\beta(W)$ is the kinematic factor related to the c.m.s. momentum of the
subprocess $\gamma^*(q)+p(P) \to \pi(k)+N(P')$ in the final state:
\bea{beta}
&&\vec k^2_f = \frac{W^2}{4}\left( 1-
\frac{(m_N\!+\!m_\pi)^2}{W^2}\right) \left( 1-
\frac{(m_N\!-\!m_\pi)^2}{W^2}\right),
\nonumber\\
&&\beta(W)=\frac{2|\vec k_f|}{W}\,.
\eea
Alternatively, instead of the polar angle dependence,  one could  use the Mandelstam $t$-variable of the $\gamma^*p \to \pi N$ subprocess
$t=(P'-P)^2$:\\[-3mm]
\beq{dt}
dt=2|\vec k_i||\vec k_f|d(\cos \theta) \,,
\eeq
where $\vec k_i$ is the c.m.s. momentum in the initial state:
\beq{cmsi}
\vec k^2_i = \frac{W^2}{4}\left( 1-
2\frac{m_N^2-Q^2}{W^2} + \frac{(m_N^2+Q^2)^2}{W^4}\right).
\eeq

Traditionally one writes the electron scattering cross section in (\ref{labsec}) in terms of the
scattering cross section for the virtual photon
\begin{equation}
 \frac{d\sigma}{dE' d \Omega'} = \Gamma_t \, d\sigma_{\gamma^\ast}\,,
\label{sigmagamma0}
\end{equation}
where
\begin{equation}
 \Gamma_t = \frac{\alpha_{\rm em}}{(2\pi)^2} \frac{W^2-m_N^2}{m_N Q^2}\frac{E'}{E}\frac{1}{1-\epsilon}
\end{equation}
is the virtual photon flux and
\begin{equation}
 \epsilon = \frac{2(1-y-m_N^2 Q^2/(s-m_N^2)^2)}{1+(1-y)^2 + 2m_N^2 Q^2/(s-m_N^2)^2}.
\end{equation}
In turn, it is convenient to separate an overall kinematic factor
in the virtual photon cross section
\begin{eqnarray}
 d\sigma_{\gamma^\ast} = \frac{\alpha_{\rm em}}{8\pi} \frac{k_f}{W} \frac{d\Omega_\pi}{W^2-m_N^2} |{\mathcal M}_{\gamma^*}|^2.
\label{sigmagamma1}
\end{eqnarray}
{}For unpolarised target $|{\mathcal M}_{\gamma^*}|^2$ can be written as a sum of contributions
\begin{eqnarray}
 |{\mathcal M}_{\gamma^*}|^2
&=& M_T
   + \epsilon \, M_L
   + \sqrt{2\epsilon(1+\epsilon)}\,M_{LT}\,\cos(\phi_\pi)
\nonumber\\&&{}
   + \epsilon M_{TT}\,\cos(2\phi_\pi)
\nonumber\\&&{}
   + \lambda \sqrt{2\epsilon(1-\epsilon)}\,M'_{LT}\, \sin(\phi_\pi)\,.
\label{sigmagamma2}
\end{eqnarray}
We will also use the notation
\begin{eqnarray}
  d\sigma^{\gamma^\ast}_{T,L,\ldots} = \frac{\alpha_{\rm em}}{8\pi} \frac{k_f}{W} \frac{d\Omega_\pi}{W^2-m_N^2} M_{T,L,\ldots}
\label{sigmagamma3}
\end{eqnarray}
for the corresponding partial cross sections.
The invariant functions $M_T$ etc. depend on the invariants of the
$\gamma^*p \to \pi N$ subprocess only; in the last term in (\ref{sigmagamma2}) $\lambda$ is the beam helicity.

\section{Generalised form factors}
\setcounter{equation}{0}
Pion electroproduction at threshold from a proton target
can be described in terms of two generalised form factors~\cite{Braun:2006td} in full analogy with the
electroproduction of a spin-1/2 nucleon resonance:
\begin{widetext}
\bea{def:G12}
 \langle N(P')\pi(k) |j_\mu^{em}(0)| p(P)\rangle
  &=&
- \frac{i}{f_\pi} \bar N(P')\gamma_5
  \left\{\left(\gamma_\mu q^2 - q_\mu \!\not\! q\right) \frac{1}{m_N^2} G_1^{\pi N}(Q^2)
    - \frac{i \sigma_{\mu\nu}q^\nu}{2m_N} G_2^{\pi N}(Q^2)\right\}N(P)\,.
\nonumber
\eea
The form factors $G_1^{\pi N}(Q^2)$ and $G_2^{\pi N}(Q^2)$ are real functions of the momentum transfer
and can be related to the S-wave transverse $E_{0+}$ and longitudinal $L_{0+}$ multipoles:
\begin{eqnarray}
E_{0+}^{\pi N}&=&\frac{\sqrt{4\pi \alpha_{\rm em}}}{8\pi f_{\pi}}
\sqrt{\frac{(2m_N+m_{\pi})^2+Q^2}{m_N^3(m_N+m_\pi)^3}}\left(Q^2 G_1^{\pi N}-\frac12 {m_N m_\pi} G_2^{\pi N}\right),
\nonumber\\
L_{0+}^{\pi N}&=&\frac{\sqrt{4\pi \alpha_{\rm em}}}{8\pi f_{\pi}}
\frac{m_N|\omega^{\rm th}_\gamma|}{2}
\sqrt{\frac{(2m_N+m_\pi)^2+Q^2}{m_N^3(m_N+m_{\pi})^3}}\left(G_2^{\pi
N}+\frac{2m_\pi}{m_N}G_1^{\pi N}\right).
\label{waves}
\end{eqnarray}
\end{widetext}
Here $\omega^{\rm th}_\gamma = (m_\pi (2m_N+m_\pi)-Q^2)/(2(m_N+m_\pi))$ is the photon energy in the c.m. frame
(at threshold). For physical pion mass both form factors are finite at $Q^2=0$. However, $G_1^{\pi^+ n}(Q^2)$
develops a singularity $\sim 1/Q^2$ at $Q^2\to 0$ in the chiral limit $m_\pi=0$.
The differential cross section at threshold is given by
\begin{equation}
 \frac{d\sigma_{\gamma^*}}{d\Omega_\pi}\Big|_{\rm th} = \frac{2|\vec{k}_f| W}{W^2-m^2}
\Big[  (E^{\pi N}_{0+})^2 + \epsilon \frac{Q^2}{(\omega^{\rm th}_\gamma)^2}
(L^{\pi N}_{0+})^2\Big].
\label{sigma_th}
\end{equation}

The LET  \cite{KR,Nambu:1997wa,Nambu:1997wb} can be formulated for the form factors directly; the corresponding
expressions can be read e.g. from Ref.~\cite{Scherer:1991cy}. Neglecting all pion mass corrections
one obtains
\begin{eqnarray}
 \frac{Q^2}{m_N^2} G_1^{\pi^0 p} &=& \frac{g_A}{2}\frac{Q^2}{(Q^2+2m_N^2)} G_M^p\,,
\nonumber\\
 G_2^{\pi^0 p} &=&    \frac{2 g_A m_N^2}{(Q^2+2m_N^2)} G_E^p\,,
\nonumber\\
  \frac{Q^2}{m_N^2} G_1^{\pi^+ n} &=& \frac{g_A}{\sqrt{2}} \frac{Q^2}{(Q^2+2m_N^2)} G_M^n + \frac{1}{\sqrt{2}}G_A\,,
\nonumber\\
 G_2^{\pi^+ n} &=&   \frac{2\sqrt{2} g_A m_N^2}{(Q^2+2m_N^2)} G_E^n\,,
\label{LET}
\end{eqnarray}
where $G^{p}_{M,E}(Q^2)$ and $G^{n}_{M,E}(Q^2)$ are the Sachs electromagnetic form factors of the proton
and neutron, respectively, and
$G_A(Q^2)$ the axial form factor induced by the charged current; $g_A\simeq 1.267$ is the axial coupling.
In this expression the terms in $G_M$ and $G_E$ correspond to the pion emission from the initial state
whereas the contribution of $G_A$ (Kroll-Ruderman term \cite {KR}) is due to the chiral rotation
of the electromagnetic current.
The correspondence between $G_1, G_2$ and $E_{0+},L_{0+}$ becomes especially simple to this accuracy:
\begin{eqnarray}
   E^{\pi N}_{0+} &=& \frac{\sqrt{4\pi \alpha_{\rm em}}}{8\pi}\frac{Q^2\sqrt{Q^2+4m^2}}{m^3 f_\pi}G_1^{\pi N}\,,
\nonumber\\
   L^{\pi N}_{0+} &=& \frac{\sqrt{4\pi \alpha_{\rm em}}}{32\pi}\frac{Q^2\sqrt{Q^2+4m^2}}{m^3 f_\pi}G_2^{\pi N}\,.
\label{waves1}
\end{eqnarray}
In the photoproduction limit $Q^2\to 0$ one obtains $E^{\pi^+ n}_{0+} \sim g_A$ and $E^{\pi^0 p}_{0+} \to 0$ so that
many more $\pi^+$ are produced at threshold compared to $\pi^0$, in agreement with experiment.

As already mentioned, although LET were applied historically to small momentum transfers $Q^2 < 0.1$ GeV$^2$
their traditional derivation using PCAC and current algebra does not seem to be affected as long as
the emitted pion remains 'soft' with respect to the initial state nucleon.
Qualitatively, one expects from (\ref{LET}) that the $\pi^0$ production cross section increases rapidly
with $Q^2$ whereas the $\pi^+$ cross section, on the contrary, decreases since contributions of $G_A$ and $G_M^n$
have opposite sign. We are not aware of any dedicated analysis of the threshold pion production data
in the $Q^2\sim 1$~GeV$^2$ region, however. Such a study can be done, e.g., in the framework of global
partial wave analysis (PWA) of $\gamma N $ and $\gamma^* N$ scattering
(cf. \cite{Drechsel:1998hk,Arndt:2002xv,Arndt:2006ym,Drechsel:2007if}) and to our opinion is long overdue.

\begin{figure}[ht]
\centerline{\epsfxsize5cm\epsffile{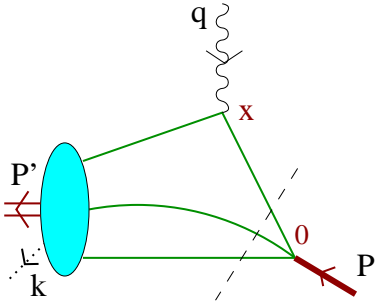}}
\caption{\small
Schematic structure of the light-cone sum rule for pion electroproduction.
}
\label{figsum}
\end{figure}

{} For $m_\pi Q^2/m_N^3 = {\mathcal O}(1)$ the LET break down: the initial state pion radiation occurs
at time scales of order $1/m_N$ rather than $1/m_\pi$ necessitating to add contributions of all
hadronic intermediate states other than the nucleon. In perturbative QCD one expects that both form factors
scale as $Q^{-6}$ at asymptotically large momentum transfers. In particular $G_1(Q^2)$ is calculable in terms of pion-nucleon 
distribution amplitudes using collinear factorisation \cite{PPS01}. 
In Ref.~\cite{Braun:2006td} we have suggested to
calculate the form factors $G_1(Q^2)$ and $G_2(Q^2)$ using the LCSR. The motivation and the
theoretical foundations of this approach are explained in \cite{Braun:2006td} and do not need to be repeated here.
The starting point is the correlation function
$$\int\! dx\, e^{-iqx}\langle N(P')\pi(k)| T \{ j^{\rm em}_\mu(x) \eta (0)\} | 0 \rangle\,, $$
where $\eta$ is a suitable operator with nucleon quantum numbers, 
see a schematic representation in Fig.~\ref{figsum}.
 When both  the momentum transfer  $Q^2$ and the momentum $P^2 = (P'-q+k)^2$ flowing in the $\eta$ vertex are large and negative, the 
 main contribution to the integral comes from the light-cone region $x^2\to 0$ and the correlation function can be
 expanded in powers of the deviation from the light cone. The coefficients in this expansion are calculable
in QCD perturbation theory and the remaining matrix elements can be identified with  pion-nucleon distribution amplitudes (DAs).
Using chiral symmetry and current algebra  these matrix elements can be reduced to the usual nucleon DAs.
On the other hand, one can represent the answer in form of the dispersion integral in $P^2$ and define the nucleon contribution
by the cutoff in the invariant mass of the three-quark system, the so-called interval of duality $s_0$ (or continuum threshold).
This cutoff does not allow large momenta to flow through the
 $\eta$-vertex so that the particular contribution shown in Fig.~\ref{figsum} is suppressed if $Q^2$ becomes too large.
Hence the large photon momentum has to find another way avoiding the nucleon vertex, which can be achieved by exchanging
gluons with large transverse momentum between the quarks. In this way the standard pQCD factorisation
arises: leading pQCD contributions correspond to three-loop $\alpha_s^2$ corrections in the LCSR approach.
For not so large $Q^2$, however, the triangle diagram in Fig.~\ref{figsum} actually dominates by the simple
reason that each hard gluon exchange involves a small  $\alpha_s/\pi \sim 0.1$ factor which is a
standard perturbation theory penalty for each extra loop.

The LCSR for pion electroproduction involve a subtlety related to the contribution of
semi-disconnected pion-nucleon contributions in the dispersion relation. In Ref.~\cite{Braun:2006td}
such contributions were neglected, the price being that the predictions could only be made
for large momentum transfers of order $Q^2 \ge 7$~GeV$^2$. For the purpose of this paper
we have reanalysed the sum rules derived in \cite{Braun:2006td} taking into account the
semi-disconnected pion-nucleon contributions explicitly, see Appendix A.
We demonstrate that, with this modification, the sum rules can be extended to
the lower $Q^2$ region so that the LET expressions in (\ref{LET}) are indeed
reproduced at $Q^2\sim 1$~GeV$^2$ to the required accuracy $\mathcal{O}(m_\pi)$.

Note that the LCSR calculation is done in the chiral limit, we do not address finite pion mass
corrections in this study. Beyond this, accurate quantitative predictions
are difficult for several reasons, e.g. because the nucleon distribution amplitudes
are poorly known. In order to minimize the dependence of various parameters in this work we only use
the LCSR to predict certain form factor ratios and then normalise to the electromagnetic nucleon form factors
as measured in experiment, see Appendix A for the details.

The sum rules in \cite{Braun:2006td} have been derived for the proton target but can easily be generalised for the neutron
as well, which only involves small modifications. We have done the corresponding analysis and
calculated the generalised form factors for the threshold pion electroproduction both from the proton,
$\gamma^* p \to \pi^0 p$, $\gamma^* p \to \pi^+ n$ and the neutron, $\gamma^* n \to \pi^0 n$, $\gamma^* n \to \pi^- p$.
The results are shown in Fig.~\ref{fig:G12proton} and Fig.~\ref{fig:G12neutron}, respectively.

\begin{figure*}[ht]
  \includegraphics[width=0.85\textwidth,angle=0]{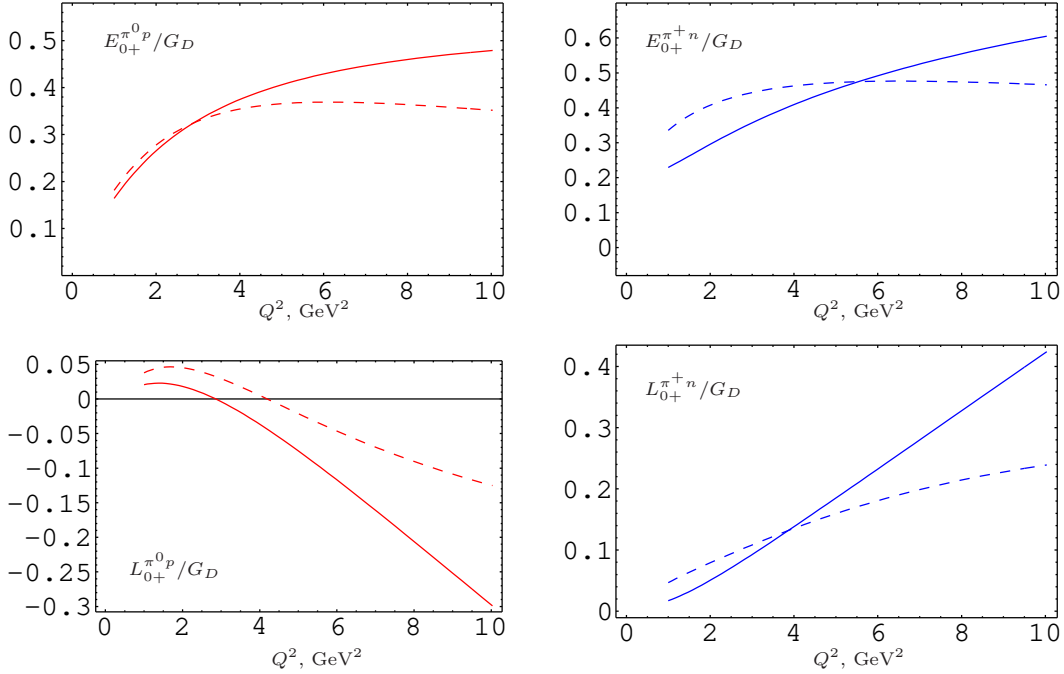}
\caption{The LCSR-based model (solid curves) for the $Q^2$ dependence of the electric and longitudinal
partial waves at threshold $E_{0+}$ and $L_{0+}$, (\ref{waves}), in units of GeV$^{-1}$,
normalised to the dipole formula (\ref{GD}).
}
\label{fig:S0+L0+proton}
\end{figure*}

The resulting LCSR-based prediction for the S-wave multipoles for the proton target
is shown by the solid curves in Fig.~\ref{fig:S0+L0+proton}.
The four partial waves at
threshold that are related to the generalised form factors through the Eq.~(\ref{waves1})
are plotted as a function of $Q^2$, normalised to the dipole formula
\begin{equation}
   G_D(Q^2) = 1/(1+Q^2/\mu_0^2)^2,
\label{GD}
\end{equation}
where $\mu_0^2 = 0.71$ GeV$^2$.
This model is used in the numerical analysis presented below. It is rather crude
but can be improved in future by calculation of radiative corrections to the sum rules and
if lattice calculations of the parameters of nucleon DAs become available.
To give a rough idea about possible uncertainties,
the ``pure'' LCSR predictions (all form factors and other input taken from the sum rules)
are shown  by dashed curves for comparison.

\section{Moving away from threshold}
\setcounter{equation}{0}

We have argued that the S-wave contributions to the threshold pion electroproduction 
are expected to deviate at large momentum transfers from the corresponding 
predictions of LET and suggested a QCD model that should be applicable in the intermediate 
$Q^2$ region. In contrast, we expect that the P-wave contributions for all $Q^2$ are 
dominated in the $m_\pi\to 0$  limit by the pion emission from the final state nucleon
(see also \cite{PPS01}). Adding this contribution, we obtain a simple expression for the      
amplitude of pion production close to threshold, $|k_f| \le m_\pi$:
\begin{widetext}
\bea{def:amp}
 \langle N(P')\pi(k) |j_\mu^{em}(0)| p(P)\rangle &=& - \frac{i}{f_\pi} \bar N(P')\gamma_5
  \left\{\left(\gamma_\mu q^2 - q_\mu \!\not\! q\right) \frac{1}{m_N^2} G_1^{\pi N}(Q^2)
    - \frac{i \sigma_{\mu\nu}q^\nu}{2m_N} G_2^{\pi N}(Q^2)\right\}N(P)\,
\nonumber\\&&
\hspace*{-3cm}{}+\frac{i c_\pi g_A}{2f_\pi[(P'+k)^2)-m_N^2]}\bar N(P')\not\! k \,\gamma_5(\not\!P'+m_N)
  \left\{F_1^p(Q^2)\left(\gamma_\mu-\frac{q_\mu\!\not\! q}{q^2}\right)+ \frac{i\sigma_{\mu\nu} q^\nu}{2m_N}F_2^p(Q^2)\right\}N(P)\,.
\nonumber\\
\eea
\end{widetext}
Hereafter $F_1^p(Q^2)$ and $F_2^p(Q^2)$ are the Dirac and Pauli electromagnetic form factors of the proton,
$c_{\pi^0} = 1$ and $c_{\pi^+} = \sqrt{2}$ are the isospin coefficients.

The separation of the generalised form factor contribution and the final state emission
in (\ref{def:amp}) can be justified in the chiral limit $m_\pi \to 0$ but involves ambiguities in contributions  $\sim \mathcal{O}(m_\pi)$.
We have chosen not to include the term $\sim \not\!k$ in the numerator of the proton propagator in the second line
in (\ref{def:amp}) so that this contribution strictly vanishes at the threshold. In addition, we found it convenient to include
the term $\sim q_\mu\!\!\not\!\! q /q^2 $ in the Lorentz structure that accompanies the $F_1$ form factor in order to make the
amplitude formally gauge invariant. To avoid misunderstanding, note that our expression is not suitable for making a
transition to the photoproduction limit $Q^2=0$ in which case, e.g. pion radiation from the initial state has to be taken
in the same approximation to maintain gauge invariance.

The amplitude in Eq.~(\ref{def:amp}) does not take into account final state interactions (FSI) which can, however, be included 
in the standard approach based on unitarity (Watson theorem), writing
(cf. e.g. \cite{Drechsel:1998hk})
\begin{equation}
 G_{1,2}^{\pi N}(Q^2) \to G_{1,2}^{\pi N}(Q^2,W) = G_{1,2}^{\pi N}(Q^2) [1+i \, t_{\pi N}]\,,
\end{equation}
where $t_{\pi N} = [\eta \exp(i\delta_{\pi N})-1]/(2i)$ is the pion-nucleon elastic scattering amplitude (for a given isospin channel)
with the S-wave phase shift $\delta_{\pi N}$ and inelasticity parameter $\eta$.
We leave this task for future, but write all expressions  for the differential cross sections and the structure 
functions for generic complex $G_1^{\pi N}$ and $G_2^{\pi N}$ so that the FSI can eventually be incorporated.
Of course, FSI in P-wave also have to be added.

Using Eq.~(\ref{def:amp}) one can calculate the differential virtual photon cross section (\ref{sigmagamma1}),
(\ref{sigmagamma2}).
The complete expressions for the invariant functions $M_{T,L,\ldots}$ are rather cumbersome but are simplified significantly
in the chiral limit $m_\pi \to 0 $ and assuming $k_f ={\mathcal O}(m_\pi)$.  We obtain
\begin{widetext}
\begin{eqnarray}
f_\pi^2 M_T &=&
\frac{4 \vec{k}_i^2 Q^2}{m_N^2} |G_1^{\pi N}|^2
 + \frac{c_\pi^2 g_A^2 \vec{k}_f^2 }{(W^2-m_N^2)^2} Q^2 m_N^2 G_M^2
   + \cos\theta \frac{c_\pi g_A |k_i| |k_f|}{W^2-m_N^2} 4 Q^2 G_M {\mathrm Re}\, G_1^{\pi N}\,,
\nonumber\\
f_\pi^2 M_L &=&
\vec{k}_i^2 |G_2^{\pi N}|^2 +
\frac{4 c_\pi^2 g_A^2 \vec{k}_f^2}{(W^2-m_N^2)^2} m^4_N G_E^2
   - \cos\theta \frac{c_\pi g_A |k_i| |k_f|}{W^2-m_N^2} 4 m_N^2 G_E {\mathrm Re}\, G_2^{\pi N} \,,
\nonumber\\
f_\pi^2 M_{LT} &=& -\sin\theta \frac{c_\pi g_A |k_i| |k_f|}{W^2-m_N^2} Qm_N
 \Big[ G_M {\mathrm Re}\,G_2^{\pi N} +  4 G_E {\mathrm Re}\,G_1^{\pi N}\Big]\,,
\nonumber\\
f_\pi^2 M_{TT} &=&0\,,
\nonumber\\
f_\pi^2 M'_{LT} &=& -\sin\theta \frac{c_\pi g_A |k_i| |k_f|}{W^2-m_N^2} Q m_N
 \Big[G_M {\mathrm Im}\,G_2^{\pi N} - 4 G_E {\mathrm Im}\,G_1^{\pi N}\Big]\,.
\end{eqnarray}
\end{widetext}
The measurements of the differential cross sections at large $Q^2$ in the threshold region would be very interesting as the
angular dependence discriminates between contributions of different origin.
In our approximation $M_{TT}=0$ (exactly) which is because we do not take into account the D-wave. Consequently, to our accuracy
the $\sim\cos(2\phi)$ contribution to the cross section is absent so that its measurement provides one with a quantitative
estimate of the importance of  the D-wave terms in the considered $W$ range.
Also note that the single spin asymmetry contribution $\sim M'_{LT}$ involves imaginary parts of the
generalised form factors that arise because of the FSI (and are calculable, at least in principle). 
The numerical results shown below are obtained
using exact expressions for $M_{T,L,\ldots}$; the difference is less than 20\% in most cases. 
Strictly speaking, this difference is beyond our accuracy although one might argue that kinematic factors in 
the calculation of the cross section should be treated exactly.

\begin{figure}[ht]
  \includegraphics[width=0.40\textwidth,angle=0]{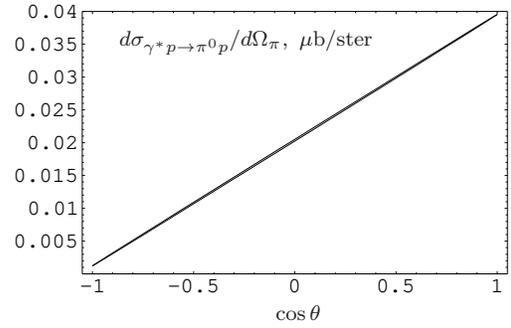}
\caption{The differential cross section $d\sigma_{\gamma^*p \to \pi^0 p}/d\Omega_\pi$ (in $\mu b$ )
as a function of $\cos\theta$ for $\phi_\pi = 135 $~grad 
for $Q^2=4.2$~GeV$^2$ and $W=1.11$~GeV.}
\label{fig:diffsigma}
\end{figure}

As an example we plot in Fig.~\ref{fig:diffsigma} the differential cross section
$d\sigma_{\gamma^*p \to \pi^0 p}/d\Omega_\pi$ [see Eq.~(\ref{sigmagamma}),(\ref{sigmagamma1})]
as a function of $\cos\theta$ for $\phi_\pi = 135^\circ$(solid curve) 
for $Q^2=4.2$~GeV$^2$ and $W=1.11$~GeV.
In fact the curve appears to be practically linear and there is no azimuthal angle dependence.
This feature is rather accidental and due to an almost complete cancellation of the contributions to $M_{LT}$ from $G_1$ and $G_2$
for the chosen value of $Q^2$. It is very sensitive to the particular choice of model parameters
and does not hold in the general case.  

\begin{figure}[ht]
  \includegraphics[width=0.40\textwidth,angle=0]{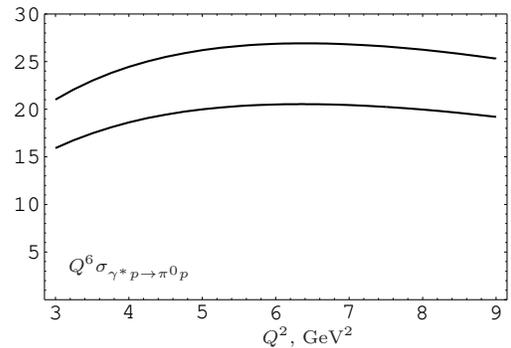}
\caption{The integrated cross section $Q^6 \sigma_{\gamma^*p \to \pi^0 p}$ (in units of $\mu b$$\times$GeV$^6$)
as a function of $Q^2$ for $W=1.11$~GeV (lower curve) and $W=1.15$~GeV (upper curve).
}
\label{fig:sigmagamma}
\end{figure}
The integrated cross section $Q^6 \sigma_{\gamma^*p \to \pi^0 p}$ (in units of $\mu b$$\times$GeV$^6$)
as a function of $Q^2$ for $W=1.11$~GeV (lower curve) and $W=1.15$~GeV (upper curve) is shown in Fig.~\ref{fig:sigmagamma}.
The predicted scaling behaviour $$\sigma_{\gamma^*p \to \pi^0 p}\sim 1/Q^6$$ 
is consistent with the SLAC measurements of the deep-inelastic structure functions \cite{Bosted:1993cc} in the 
threshold region that we are going to discuss next.

To avoid misunderstanding we stress that the estimates of the cross sections presented here
are not state-of-the-art and are only meant to provide one with the order-of-magnitude estimates
of the threshold cross sections that are to our opinion most interesting.
These estimates can be improved in many ways, for example taking into account
the energy dependence of the generalised form factors generated by the FSI
and adding a model for the D-wave contributions. The model can also be tuned to reproduce the
existing lower $Q^2$ and/or larger $W$ experimental data.
A more systematic approach could be to study the threshold production in the framework of global
PWA of $\pi N $ and $\gamma^* N$ scattering using QCD-motivated S- and P-wave multipoles and the  
D- and higher partial waves estimated from the analysis of the resonance region  
(cf. \cite{Drechsel:1998hk,Arndt:2002xv,Arndt:2006ym,Drechsel:2007if}) where there is high statistics.

\section{Structure Functions}
\setcounter{equation}{0}

The deep-inelastic structure functions $F_1(W,Q^2)$ and  $F_2(W,Q^2)$
are directly related to the total cross section of the virtual photon--proton
interaction. For the longitudinal photon polarisation one obtains
\beq{sigmaL}
\sigma^{\gamma^*}_L=\frac{8\pi^2
\alpha_{\rm em}}{W^2-m_N^2}\left(\frac{1+4x_B^2m_N^2/Q^2}{2 x_B}F_2-F_1\right)
\eeq
and for the transverse
\beq{sigmaT}
\sigma^{\gamma^*}_T=\frac{8\pi^2
\alpha_{\rm em}}{W^2-m_N^2}F_1\,.
\eeq
Here we introduced the Bjorken variable 
$$x_B=Q^2/(2 P\cdot q)=Q^2/(W^2+Q^2-m^2_N)).$$
It is customary to write the total cross section
$\sigma^{\gamma^*} = \sigma^{\gamma^*}_T + \epsilon\sigma^{\gamma^*}_L$
in terms of the structure function $F_2(W,Q^2)$ and $R= \sigma^{\gamma^*}_L/ \sigma^{\gamma^*}_T$,
the ratio of the longitudinal to transverse cross sections:
\bea{sigmagamma}
\sigma^{\gamma^*}&=&
\frac{4\pi^2 \alpha_{\rm em}\left(1+4x_B^2m_N^2/Q^2\right)}{x_B(W^2-m_N^2)}
F_2(W,Q^2)
\nonumber\\&&{}\times\left(1-(1-\epsilon)
\frac{R}{1+R}\right).
\eea
In the threshold region $x_B\to 1$, $W-m_N-m_\pi \sim \mathcal{O}(m_\pi)$, the structure
functions can be calculated starting from the amplitude in Eq.~(\ref{def:amp}).
In particular for $F_2(W,Q^2)$ we obtain
\begin{widetext}
\bea{F2exact} \hspace*{-1cm}
F_2(W,Q^2)&=&\frac{\beta(W)}{(4\pi
f_\pi)^2}(W^2+Q^2-m_N^2)(W^2+m_N^2-m_\pi^2)
\nonumber\\&&\hspace*{-1cm}\times \sum_{\pi^0,\pi^+}
\Bigg\{\frac{1}{2 m_N^4 W^2} \left(|Q^2G_1^{\pi N}|^2 + \frac14
m_N^2 Q^2 |G_2^{\pi N}|^2 \right)
 + \frac{c_\pi^2 g_A^2\beta^2(W) W^2}{8(W^2-m_N^2)^2}\left( (F_1^p)^2 + \frac{Q^2}{4 m_N^2} (F_2^p)^2\right)
\nonumber\\&&{}
 -\frac{c_\pi g_A \beta^2(W) Q^2 W^2}{2 m_N^2 (W^2-m_N^2)(W^2+m_N^2-m_\pi^2)}
{\mathrm Re} \left( F_1^p G_1^{\pi N} +\frac{1}{4} F_2^p G_2^{\pi N}\right)
\Bigg\}.
\eea
Similar to the differential cross sections, expressions for the structure functions are
simplified considerably in the chiral limit $m_\pi \to 0 $ and assuming $k_f ={\mathcal O}(m_\pi)$:
we have to retain the kinematic factor $W^2 \beta^2(W) = 4 |\vec{k}_f|^2$ but can neglect the
pion mass corrections and the difference $W^2-m_N^2$ whenever possible. The results are
\begin{eqnarray}
F_1(W,Q^2)&=&\frac{\beta(W)}{(4\pi f_\pi)^2}
\sum_{\pi^0,\pi^+}
\Bigg\{\frac{Q^2+4 m_N^2}{2m_N^4} |Q^2 G_1^{\pi N}|^2
+\frac{c_\pi^2 g_A^2 W^2 \beta^2(W)}{8 (W^2-m_N^2)^2} Q^2 m_N^2 G_M^2\Bigg\},
\nonumber\\
F_2(W,Q^2)&=&\frac{\beta(W)}{(4\pi f_\pi)^2}\!
\!\sum_{\pi^0,\pi^+}\!\!\!
\Bigg\{\frac{Q^2}{m_N^4}\!
\left(|Q^2G_1^{\pi N}|^2 \!+\! \frac14 m_N^2 Q^2 |G_2^{\pi N}|^2 \right)
 \!+\! \frac{c_\pi^2 g_A^2 W^2 \beta^2(W) Q^2 m_N^2}{4(W^2-m_N^2)^2}
\!\!\left(\frac{Q^2 G_M^2 \!+\! 4 m_N^2 G_E^2}{Q^2+ 4 m_N^2}\right)\!\!\!\Bigg\},
\nonumber\\
g_1(W,Q^2)&=&\frac{\beta(W)}{(4\pi f_\pi)^2}
\sum_{\pi^0,\pi^+}
\Bigg\{\frac{Q^2}{2m_N^4}\Big[|Q^2G_1^{\pi N}|^2  - m_N^2{\mathrm Re}(Q^2 G_1^{\pi N}G_2^{\ast,\pi N})\Big]
 + \frac{c_\pi^2 g_A^2 W^2\beta^2(W)}{8(W^2-m_N^2)^2}Q^2 m_N^2 G_M F^p_1\Bigg\},
\nonumber\\
g_2(W,Q^2)&=&-\frac{\beta(W)}{(4\pi f_\pi)^2}\!\!
\sum_{\pi^0,\pi^+}\!\!
\Bigg\{
 \frac{Q^2}{2m_N^4}\Big[|Q^2 G_1^{\pi N}|^2 + \frac{1}{4}Q^2{\mathrm Re}(Q^2 G_1^{\pi N}G_2^{\ast,\pi N}) \Big]
+ \frac{c_\pi^2 g_A^2 W^2 \beta^2(W)}{32(W^2-m_N^2)^2}Q^4 G_M F^p_2
\Bigg\},
\label{sfsimple}
\end{eqnarray}
\end{widetext}
\noindent
where, for completeness, we included the polarised structure functions $g_1(W,Q^2)$ and $g_2(W,Q^2)$.
Note that in this limit the contributions $\sim |G_{1,2}^{\pi N}|^2$ and  $\sim |G_{E.M}^{p}|^2$ 
can be identified with the pure S-wave and P-wave, respectively.
Numerically, the difference between the complete expressions like the one in (\ref{F2exact}) and the ones in 
the chiral limit $m_\pi \to 0 $ in
(\ref{sfsimple}) is less than 20\% and, strictly speaking, beyond our accuracy.

\begin{figure}[ht]
  \includegraphics[width=0.40\textwidth,angle=0]{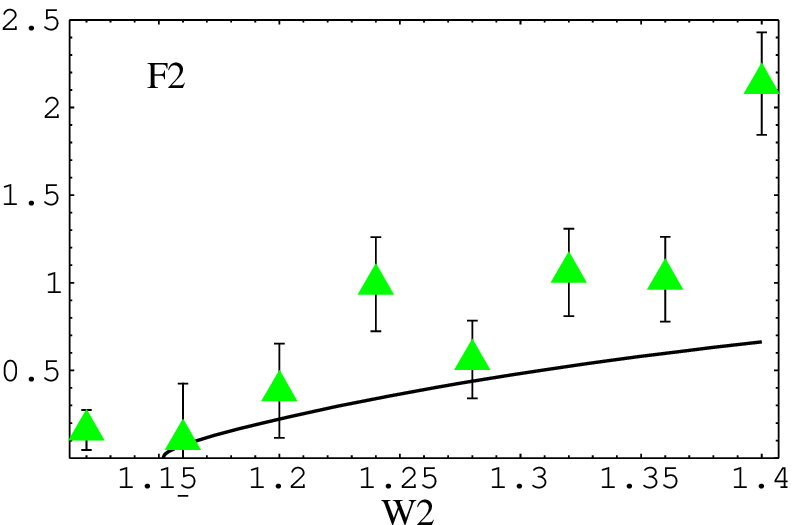}\\[2mm]
  \includegraphics[width=0.40\textwidth,angle=0]{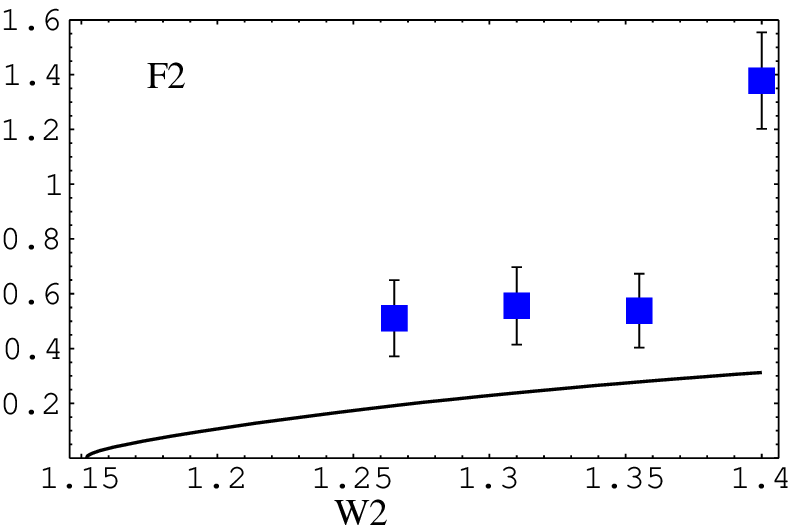}
\caption{The structure function $F^p_2(W,Q^2)$ as a function of $W^2$ scaled by a factor
 $10^3$ compared to the SLAC E136 data \cite{Bosted:1993cc} at the average value $Q^2 = 7.14$~GeV$^2$ (upper panel) and
$Q^2 = 9.43$~GeV$^2$ (lower panel).
}
\label{fig:E136}
\end{figure}

With these expressions at hand, one can easily obtain the longitudinal to transverse cross section
ratio. In particular, at the threshold we get, in the $m_\pi\to 0$ limit,
\beq{Rth}
R_{\rm th}= \lim_{W\to W_{\rm th}}R=\left(\frac{m_N\,G_2^{\pi N}}{2 Q\,G_1^{\pi N}}\right)^2.
\eeq
In the pQCD regime $Q^2\to\infty$ one expects that $G_2^{\pi N}$ is suppressed compared to $Q^2 G_1^{\pi N}$ by a power
of $1/Q^2$ and thus $R_{\rm th}$ scales like $R_{\rm th}\sim 1/Q^2$, same as in the deep-inelastic region;
this scaling behavior was assumed in the analysis of the experimental data in \cite{Bosted:1993cc}.
In the LCSR approach the $Q^2$ dependence of $G_1^{\pi N}$ and $G_2^{\pi N}$ turns out to be
similar to that of the proton Dirac, $F_1^p$, and
Pauli, $F_2^p$, electromagnetic form factors, respectively. Since in the intermediate $Q^2$ range
$1 < Q^2 < 6$ GeV$^2$ the Pauli form factor decreases more slowly compared to the pQCD counting rules
and the observed suppression is rather $F_2/F_1 \sim 1/Q$ instead of expected $1/Q^2$, the
$R_{\rm th}$ ratio is enhanced. With our parameterisation of the form factors one
obtains using soft pion limit result in Eq.~(\ref{Rth}) that $R_{\rm
th}=0.21$ and is independent on $Q^2$. The complete expressions for the amplitudes
give a somewhat smaller value  $R_{\rm th}=0.13\div 0.16$ for $Q^2=4\div 9$ GeV$^2$, with
a weak $Q^2$ dependence.

The comparison of the LCSR-based  predictions for the structure function $F^p_2(W,Q^2)$ in the threshold region 
$W^2<1.4$~GeV$^2$ to the SLAC E136 data \cite{Bosted:1993cc} at the average value $Q^2 = 7.14$~GeV$^2$ and 
$Q^2 = 9.43$~GeV$^2$ is shown in Fig.~\ref{fig:E136}.
The predictions are generally somewhat below these data ($\sim 30-50$\%), apart from the last data point at $W^2=1.4$~GeV$^2$
which is significantly higher.
Note that in our approximation there is no D-wave contribution and the final state
interaction is not included. Both effects can increase the cross section so that we consider the agreement as satisfactory.
We believe that the structure function at $W^2=1.4$~GeV$^2$ already contains a considerable D-wave contribution and also
one from the tail of the $\Delta$-resonance and thus cannot be compared with our model, at least in
its present form.

The results shown in  Fig.~\ref{fig:E136} are obtained using the complete expression
for the structure function $F_2$ given in Eq.~(\ref{F2exact}). The difference with using the
simplified expression in  Eq.~(\ref{sfsimple}) is, however, small. In particular the
interference contributions $\sim F_1 G_1^{\pi N}$ etc. in the third line in  Eq.~(\ref{F2exact}) do not exceed 10-15\%.

\begin{figure}[ht]
  \includegraphics[width=0.40\textwidth,angle=0]{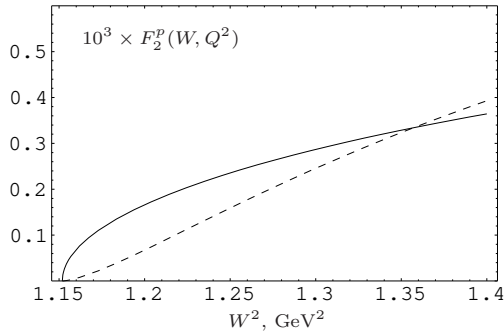}
\caption{The S-wave (solid) vs. the P-wave (dashed) contribution to the structure function $F^p_2(W,Q^2)$
as a function of $W^2$ for $Q^2 = 7.14$~GeV$^2$.
}
\label{fig:SPwaveratio}
\end{figure}

Further, in Fig.~\ref{fig:SPwaveratio} we show the  contributions of the S-wave (solid curve) and P-wave (dashed)
to the structure function $F^p_2(W,Q^2)$ separately as a function of $W^2$ for $Q^2 = 7.14$~GeV$^2$.
It is seen that the P-wave contribution is smaller than the S-wave one up to $W\sim 1.16$~GeV. 

\begin{figure}[ht]
  \includegraphics[width=0.40\textwidth,angle=0]{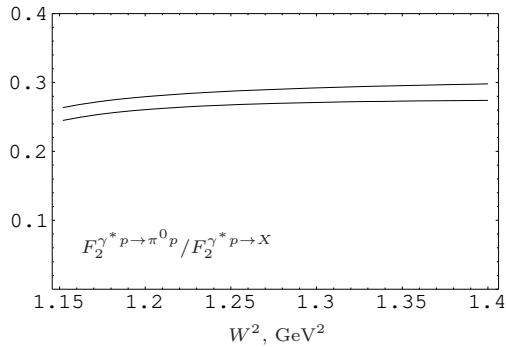}
\caption{The contribution of the $\pi^0 p$  final state to the structure function $F^p_2(W,Q^2)$
as a function of $W^2$ for $Q^2 = 3$~GeV$^2$ (upper curve) and
$Q^2 = 9$~GeV$^2$ (lower curve).
}
\label{fig:pi0piplusratio}
\end{figure}

The contribution of the $\pi^0 p$ final state to the structure function $F^p_2(W,Q^2)$ is predicted to be around 30\%
and nearly constant in a broad $Q^2$ and $W$-range, see Fig.~\ref{fig:pi0piplusratio}.

Last but not least, the ratios of the proton and the neutron structure functions in the threshold region
are of interest as a manifestation of helicity counting rules in pQCD: a quark with largest momentum fraction
of the hadron tends to carry also its helicity \cite{Farrar:1975yb}, see e.g. \cite{Brodsky:1994kg,Avakian:2007xa}
for recent applications and discussion.  
Using LCSR predictions for the generalised form factors for the pion threshold electroproduction 
from the neutron target (see Appendix A) we obtain  
for $Q^2>7$~GeV$^2$
\begin{eqnarray}
 \lim_{W\to W_{\rm th}} \frac{F_2^n(W,Q^2)}{F_2^p(W,Q^2)} &=& 0.41
 (0.23)\,,\\
 \lim_{W\to W_{\rm th}} \frac{g_1^n(W,Q^2)}{g_1^p(W,Q^2)} &=& 0.44
 (0.21)\,,
\label{pn-ratio}
\end{eqnarray}
with a very weak  dependence on $Q^2$. The numbers in parenthesis correspond to the LCSR results obtained with the asymptotic DAs.
The first ratio in (\ref{pn-ratio}) appears to be in a striking agreement with the parton model prediction
$F_2^n/F_2^p = 3/7$ \cite{Farrar:1975yb} for $x_B\to 1$, although the present approach seems to be very different.

\section{Conclusions}

The rapid development of experimental techniques is making possible to study threshold pion production
with photon virtuality in a few GeV$^2$ range. 
The physics of threshold production is very rich and interesting, and allows for better theoretical 
understanding, as compared to the more conventional resonance region, based on chiral symmetry of QCD 
in the limit of vanishing pion mass. The momentum transfer dependence of the S-wave multipoles 
is especially intriguing. For small $Q$ of the order of the pion mass it is well described 
by the chiral perturbation theory \cite{Bernard:1995dp}. The expansion in powers of $Q^2$ which is endemic to CHPT as a local 
effective theory is, however, not warranted. The derivation of classical low-energy theorems \cite{KR,Nambu:1997wa,Nambu:1997wb}
does not seem to be affected as long as $Q^2 < \Lambda^3/m_\pi$ where $\Lambda$ is a certain hadronic scale,
at least for the leading contributions in the $m_\pi \to 0$ limit. This implies, in particular, that the relation 
between the $\gamma^* p \to \pi^+ n$ amplitude and the proton axial form factor \cite{KR} holds true well beyond the 
applicability range of CHPT, say, for $Q^2\sim 1$~GeV$^2$. These expectations have to be checked, as the first task.
{}For larger $Q^2$ in a several GeV$^2$ region the LET are not expected to hold
 because the produced pion cannot remain 'soft'
to both initial and final state nucleons simultaneously. Main contribution of this work is to suggest a realistic model 
for the S-wave transverse $S_{0+}$ and longitudinal $L_{0+}$ multipoles for the intermediate $Q^2 \sim 1-10$~GeV$^2$ region,
based on chiral symmetry and light-cone sum rules. For asymptotically large $Q^2$, the $S_{0+}$ can be calculated in 
pQCD in terms of chirally rotated nucleon distribution amplitudes \cite{PPS01}.
The P-wave contributions appear to be much simpler: they
are dominated in the $m_\pi\to0$ limit by the emission from the final state and are given in terms of the 
electromagnetic nucleon form factors for all momentum transfers.  
In Section~4 we have introduced a simple model for the electroproduction close to threshold, complementing the S-wave
form factor-like contributions by the P-wave terms.
In this framework, detailed predictions are worked
out for the differential cross sections from the proton target and also for the structure functions measured in
the deep-inelastic scattering experiments.
In future we expect that the extraction from the data of the most interesting S-wave multipoles can be done 
in the framework of a global partial wave analysis, cf.  \cite{Drechsel:1998hk,Arndt:2002xv,Arndt:2006ym,Drechsel:2007if},
which have to be adapted, however, to the threshold kinematics.  

In addition to the threshold production, there exists another interesting kinematic region where the pion is 
produced backwards in the c.m. frame and is 'soft' with respect to the initial proton, 
i.e. has small momentum in the laboratory frame \cite{Lansberg:2007ec}. 
In the limit $m_\pi\to 0$ the corresponding amplitudes are given by form factor-like contributions that are very similar to the ones 
considered here, and can be estimated in the LCSR approach in terms of pion-to-nucleon transition distribution amplitudes 
introduced in \cite{Lansberg:2007ec}. In addition, one has to take into account pion emission from the initial state.
The problem is, however, that in the accessible $Q^2$ range the invariant energy of the outgoing pion-nucleon system 
appears in this case to be in the resonance region so that FSI would have to be taken into account explicitly. 
The corresponding calculation goes beyond the scope of this paper.

\section*{Acknowledgements}
\setcounter{equation}{0}
We gratefully acknowledge useful discussions with A.~Afanasev, V.~Kubarovsky, A.~Lenz, A.~Sch{\"a}fer, P.~Stoler and
I.~Strakovsky on various aspects of this project, 
and U.~Meissner for bringing Ref.~\cite{Bernard:1992ys} to our attention and useful comments.
V.B. is grateful to IPPP for hospitality and financial support
during his stay at Durham University where this work was
finalised.
The work of D.I. was partially supported by grants from
RFBR-05-02-16211, NSh-5362.2006.2 and BMBF(06RY258).
The work by A.P. was supported by the Studienstiftung des deutschen Volkes.

\appendix
\renewcommand{\theequation}{\Alph{section}.\arabic{equation}}
%

%
\begin{widetext}
%

\section*{Appendices}

\section{Light-Cone Sum Rules}
\setcounter{equation}{0}

{}For technical reasons, it is convenient to write the  sum rules for the complex conjugated amplitude
with the pion-nucleon pair in the initial state. To this end we consider the
leading twist projection of the correlation function \cite{Braun:2006td}
\bea{correlator}
z^\nu\Lambda^+T_{\nu}^{\pi N}(P,q)
 &=&  z^\nu\Lambda^+i\!\int\! \dd^4 x \, e^{i q x}
\bra{0} T\left\{\eta (0) j_{\nu}^{\mathrm{em}}(x)\right\} \ket{N(P) \pi(k)}
\nonumber\\
&=&  \frac{i}{f_\pi}(pz+kz)\gamma_5 \left\{ m_N \mathcal{A}(P'^2,Q^2) +
    \!\not\!q_\perp \mathcal{B}(P'^2,Q^2)\right\} N^+(P)\,,
\eea
where $P'= P+k-q$, $z^\mu$ is a light-like vector such that $z^2=0$ and
$q\cdot z=0$, $\Lambda^+ = (\!\not\!{p}\! \not\!{z})/(2 p\cdot z)$
is the projector on the ``plus'' components of the nucleon spinor $N_+(P)=\Lambda_+N(P)$.
Further, $p_\mu = P_\mu  - (1/2) \, z_\mu m_N^2/(P\cdot z)$, 
$q_\perp^\mu  = q_\mu -z_\mu (p\cdot q)/(p\cdot z)$ is the transverse component of the momentum transfer
and
\bea{current}
\eta_p(x) &=& \ep^{ijk} \left[u^i(x) C\ga_\mu u^j(x)\right]\,\ga_5 \ga^\mu d^k(x)\,,
\nonumber\\
\eta_n(x) &=& -\ep^{ijk} \left[d^i(x) C\ga_\mu d^j(x)\right]\,\ga_5 \ga^\mu u^k(x)
\eea
are the so-called Ioffe interpolating currents \cite{Ioffe:1981kw} for the proton and the 
neutron, respectively.
The corresponding coupling
\bea{lambda1}
\bra{0} \eta(0)\ket{N(P)}  &=& \lambda_1 m_N N(P)\, 
\eea
is the same for the proton and the neutron, $\lambda_1^p=\lambda_1^n$, because of the isospin symmetry.

The invariant functions $\mathcal{A}(P'^2,Q^2)$ and $\mathcal{B}(P'^2,Q^2)$ can be calculated in the
Euclidean region $P'^2 < 0, Q^2 < 0$ in terms of the pion-nucleon generalised distribution amplitudes
using the operator product expansion. The corresponding expressions are given in Eq.~(4.17)
in Ref.~\cite{Braun:2006td} to leading order in the QCD coupling. 
The sum rules are derived using continuum-subtracted
Borel transforms
\bea{Borel1}
     \mathbb{B}_{P'^2}[\mathcal{A}](M^2,Q^2) &=& \frac{1}{\pi}\int_0^{s_0}ds\, e^{-s/M^2}\,\Im \mathcal{A}(s,Q^2)
\eea
and similar for $\mathbb{B}_{P'^2}[\mathcal{A}](M^2,Q^2)$. The explicit expressions are \cite{Braun:2006td}
\bea{LCSRIoffe}
e^{m_N^2/M^2}\mathbb{B}_{P'^2}[\mathcal{A}](M^2,Q^2) &=&
\left[ \int_{x_0}^1 \dd x \left( - \frac{\varrho_2^a(x)}{x}  +
\frac{\varrho_4^a(x)}{x^2M^2} \right)
        \exp{ \left( - \frac{\bar x Q^2}{x M^2}  + \frac{x
m_N^2}{M^2}\right)}\,
+
\frac{\varrho_4^a(x_0)\,e^{-(s_0-m_N^2) /M^2} }{Q^2 + x_0^2 m_N^2} \right],
\nn
\\
e^{m_N^2/M^2}\mathbb{B}_{P'^2}[\mathcal{B}](M^2,Q^2) &=&
\left[ \int_{x_0}^1 \dd x \left( - \frac{\varrho_2^b(x)}{x}  +
\frac{\varrho_4^b(x)}{x^2M^2} \right)
        \exp{ \left( - \frac{\bar x Q^2}{x M^2}  + \frac{x
m_N^2}{M^2}\right)}\,
+
\frac{\varrho_4^b(x_0)\,e^{-(s_0-m_N^2) /M^2} }{Q^2 + x_0^2 m_N^2} \right],
\nonumber\\
\eea
where the factor $e^{m_N^2/M^2}$ is included for later convenience and the
spectral functions $\varrho^{a,b}_{2,4}(x)$ are given in terms of the generalised pion-nucleon distribution
amplitudes. In the notation of Ref.~\cite{Braun:2006td}
\bea{abbrev}
\varrho_2^{a}(x) & = &
       2e_d \bigg\{ \widetilde{V}_{123}^{\pi N} + x \int_0^{\bar x} dx_1 V_3^{\pi N}(x_i)
\bigg\}
  + 2 e_u \bigg\{ x \int_0^{\bar x}dx_1 \left[ -2 V_1 + 3V_3+A_3
\right]^{\pi N}(x_i)
        - \widehat{V}_{123}^{\pi N} + \widehat{A}_{123}^{\pi N} \bigg\},
\nn \\
\varrho_4^{a}(x) & = &
       2e_d \bigg\{Q^2 \widetilde{V}_{123}^{\pi N}+ x^2 m_N^2
\widetilde{V}_{43}^{\pi N}\bigg\}
+ 2 e_u  \bigg\{ Q^2 \left( \widehat{V}_{123}^{\pi N} + \widehat{A}_{123}^{\pi N} \right)
- x^2 m_N^2 \left[\widehat{V}_{1345}^{\pi N} - 2\widehat{V}_{43}^{\pi N}+ \widehat{A}_{34}^{\pi N}\right]
 \nn \\ && {}
- 2x m_N^2  \left( \mathcal{V}_1^{\pi N,M(u)} + \widehat{\widehat{V}}_{123456} ^{\pi N}\right) \bigg\},
\nn \\
\varrho_2^{b}(x) & = &
- 2 e_d \left\{ \int_0^{\bar x} dx_1 V_1^{\pi N}(x_i)    \right\}
+ 2 e_u \left\{ \int_0^{\bar x}dx_1 \left[V_1+A_1\right]^{\pi N}(x_i) \right\},
\nn \\
\varrho_4^{b}(x) & = &  -2e_d m_N^2
\bigg\{ \mathcal{V}_1^{\pi N,M(d)} - x \left[\widetilde{V}_{123} -
\widetilde{V}_{43} \right]^{\pi N} \bigg\}
  + 2 e_u m_N^2 \bigg\{ \left[\mathcal{V}_1^{\pi N,M(u)}+\mathcal{A}_1^{\pi N,M(u)}\right]
\nn \\ &&{}
  + x \left[\widehat{V}_{1345} +
\widehat{V}_{123}+\widehat{A}_{123}-2\widehat{V}_{43}+\widehat{A}_{34}\right]^{\pi N}
\bigg\}.
\eea
The sum rules are obtained matching the above expressions with the dispersion representation for the
correlation functions in terms of hadronic states below the continuum threshold.
The contributions of interest to (\ref{correlator}) are those singular in the vicinity of $P'^2 \to m_N^2$,
see Fig.~\ref{figdisp}.
\begin{figure}[ht]
\centerline{\epsfxsize14cm\epsffile{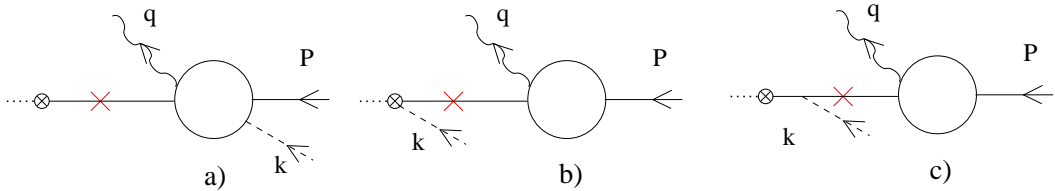}}
\caption{\small
Schematic structure of the pole terms in the correlation function
 (\ref{correlator})
}
\label{figdisp}
\end{figure}
Note that in addition to the nucleon pole, Fig.~\ref{figdisp}a, one has to take into account the
semidisconnected contribution with the pion-nucleon intermediate state. In the soft-pion limit $m_\pi\to0$
and not too far from the threshold they can be estimated as due to the chiral rotation
of the Ioffe current, Fig.~\ref{figdisp}b, and pion emission in the final state, Fig.~\ref{figdisp}c.
Taken together, these two contributions correspond to the approximation for the Ioffe current coupling to
a pion-nucleon state:
\bea{Ioffe_piN}
 \bra{0} \eta_p(0)\ket{p(P'-k)\pi^0(k)}  &=& \frac{i\lambda_1^p m_N}{2f_\pi}
\left[1-\frac{g_A}{P'^2-m_N^2}(\not\!P'-\not\!k+m_N)\not\!k\right]\gamma_5 N_p(P'-k) \,, 
\nonumber\\
\bra{0} \eta_p(0)\ket{n(P'-k)\pi^+(k)}  &=& \frac{i\lambda_1^p m_N}{\sqrt{2}f_\pi}
\left[1-\frac{g_A}{P'^2-m_N^2}(\not\!P'-\not\!k+m_N)\not\!k\right]\gamma_5 N_n(P'-k) \,,
\nonumber\\
\bra{0} \eta_n(0)\ket{n(P'-k)\pi^0(k)}  &=& \frac{-i\lambda_1^p m_N}{2f_\pi}
\left[1-\frac{g_A}{P'^2-m_N^2}(\not\!P'-\not\!k+m_N)\not\!k\right]\gamma_5 N_n(P'-k) \,, 
\nonumber\\
\bra{0} \eta_n(0)\ket{p(P'-k)\pi^-(k)}  &=& \frac{i\lambda_1^p m_N}{\sqrt{2}f_\pi}
\left[1-\frac{g_A}{P'^2-m_N^2}(\not\!P'-\not\!k+m_N)\not\!k\right]\gamma_5 N_p(P'-k) \,.
\eea
{}For the sum of the three contributions in Fig.~\ref{figdisp} to the correlation function
$T_{\nu}^{\pi^0 p}(P,q)$ one obtains, for example
\begin{eqnarray}
T_{\nu}^{\pi^0 p}(P,q)
&=& \frac{i\lambda_1^p m_N}{f_\pi} \left\{\frac{((1+\delta)\not\!P -\not\!q + m_N)\gamma_5}{m_N^2-P'^2}
 \left[(\gamma_\nu q^2 -q_\nu \not\!q) \frac{G_1^{\pi^0 p}}{m_N^2}-
\frac{i\sigma_{\nu\mu}q^\mu}{2m_N}G_2^{\pi^0 p}\right]
\right.
\nonumber\\
&&{}+ \frac{1}{2}\frac{(1+\delta) \gamma_5(\not\!P -\not\!q + m_N)}{[m_N^2(1+\delta)^2+\delta Q^2]-P'^2}
\left[\gamma_\nu F_1^{p}- \frac{i\sigma_{\nu\mu} q^\mu}{2m_N}F_2^{p}\right]
\label{rep1}
\\
&&{}\left.-\frac{1}{2[Q^2+m_N^2(2+\delta)]}
\frac{(1+\delta)g_A(\not\!P -\not\!q + m_N)\gamma_5}{[m_N^2(1+\delta)^2+\delta Q^2]-P'^2}
 \left[(\gamma_\nu q^2 -q_\nu\! \not\!q) G_M^p- \frac{i\sigma_{\nu\mu}q^\mu}{2m_N} 4m_N^2 G_E^p\right]
\right\}N(P)\,,
\nonumber
\end{eqnarray}
where $\delta=m_\pi/m_N$ and the threshold kinematics is assumed for the initial state, i.e.
$k_\mu = \delta P_\mu$.
Making the appropriate projections one obtains for the proton target, after a short calculation
\bea{projections-lhs}
 \mathcal{A}^{\pi^0 p} & = & \phantom{-} \frac{2\lambda_1^p}{m_N^2- P'^2}\,\frac{Q^2}{m_N^2} G^{\pi^0 p}_1(Q^2)
  +   \frac{\lambda_1^p}{m_N^2+\delta(2m_N^2+Q^2)-P'^2}\left[F_1^p(Q^2)-\frac{g_A Q^2}{Q^2+2m_N^2}G^p_M(Q^2)\right]\,,
\nonumber\\
 \mathcal{B}^{\pi^0 p} & = &  {-}\frac{\lambda_1^p}{m_N^2- P'^2} \, G^{\pi^0 p}_2(Q^2) +
   \frac{\lambda_{1}^p}{m_N^2+\delta(2m_N^2+Q^2)-P'^2}\left[\frac{1}{2}F^p_2(Q^2)
          +\frac{2g_A m_N^2}{Q^2+2m_N^2}G^p_E(Q^2)\right]\,,
\nonumber\\
 \mathcal{A}^{\pi^+ n} & = & \phantom{-} \frac{2\lambda_1^p}{m_N^2- P'^2}\,\frac{Q^2}{m_N^2} G^{\pi^+ n}_1(Q^2)
  +   \frac{\sqrt{2}\lambda_1^p}{m_N^2+\delta(2m_N^2+Q^2)-P'^2}\left[F^n_1(Q^2)
     -\frac{g_A Q^2}{Q^2+2m_N^2}G^n_M(Q^2)\right]\,,
\nonumber\\
 \mathcal{B}^{\pi^+ n} & = &  {-}\frac{\lambda_1^p}{m_N^2- P'^2} \, G^{\pi^+ n}_2(Q^2) +
   \frac{\sqrt{2}\lambda_1^p}{m_N^2+\delta(2m_N^2+Q^2)-P'^2}\left[\frac{1}{2}F^n_2(Q^2)
          +\frac{2g_A m_N^2}{Q^2+2m_N^2}G^n_E(Q^2)\right]\,.
\eea
Making the Borel transformation and equating the result to the QCD calculation in (\ref{LCSRIoffe}) we end up
with the sum rules
\bea{finalSR}
\frac{Q^2}{m_N^2}G^{\pi^0 p}_1 &=&
 \frac{e^{m_N^2/M^2}}{2\lambda_1^p} \mathbb{B}_{P'^2}[\mathcal{A}^{\pi^0 p}](M^2,Q^2) -\frac{1}{2}e^{-\delta(2m_N^2+Q^2)/M^2}
       \left[F_1^p(Q^2)-\frac{g_A Q^2}{Q^2+2m_N^2}G^p_M(Q^2)\right]\,,
\nonumber\\
  G^{\pi^0 p}_2 &=&
 - \frac{e^{m_N^2/M^2}}{\lambda_1^p} \mathbb{B}_{P'^2}[\mathcal{B}^{\pi^0 p}](M^2,Q^2) +e^{-\delta(2m_N^2+Q^2)/M^2}
\left[\frac{1}{2}F^p_2(Q^2) +\frac{2g_A m_N^2}{Q^2+2m_N^2}G^p_E(Q^2)\right]\,,
\nonumber\\
\frac{Q^2}{m_N^2}G^{\pi^+ n}_1 &=&
 \frac{e^{m_N^2/M^2}}{2\lambda_1^p}\mathbb{B}_{P'^2}[\mathcal{A}^{\pi^+ n}](M^2,Q^2)
    -\frac{1}{\sqrt{2}}e^{-\delta(2m_N^2+Q^2)/M^2}
    \left[F^n_1(Q^2) -\frac{g_A Q^2}{Q^2+2m_N^2}G^n_M(Q^2)\right]\,,
\nonumber\\
  G^{\pi^+ n}_2 &=& - \frac{e^{m_N^2/M^2}}{\lambda_1^p} \mathbb{B}_{P'^2}[\mathcal{B}^{\pi^+ n}](M^2\!,Q^2) + e^{-\delta(2m_N^2+Q^2)/M^2}
\!\!\left[\frac{1}{\sqrt{2}}F^n_2(Q^2) +\frac{2\sqrt{2}g_A m_N^2}{Q^2+2m_N^2}G^n_E(Q^2)\right]\!\!.
\eea
Note that the contribution of the pion-nucleon intermediate state is suppressed compared to the nucleon
one by an extra factor $\exp\{-\delta[2m_N^2+Q^2]/M^2\}$ which reflects the fact that the corresponding singularity in the
complex $P'^2$ plane is shifted by the amount $\delta(2m_N^2+Q^2)$. For momentum transfers larger than
$Q^2 \sim 7.3$~GeV$^2$ this contribution moves to the continuum region $P'^2>s_0\simeq (1.5$~GeV$)^2$ and can be dropped.
This is the limit considered in Ref.~\cite{Braun:2006td}. For small momentum transfers, on the other hand,
one can apply the current algebra techniques directly to the correlation function (\ref{correlator})
so that it can be written in terms of the correlation functions without the pion and
involving chirally-rotated currents
\begin{equation}
 T_\nu^{\pi N} (P,q) =-\frac{i}{f_\pi}\left[
 i \!\int\! d^4x\, e^{iqx}\langle 0|T\{{[Q^a_5,\eta(0)]}j_\nu^{\mathrm{em}}(x)\}|N(P)\rangle
+i \!\int\! d^4x\, e^{iqx}\langle 0|T\{\eta_p(0){[Q^a_5,j_\nu^{\mathrm{em}}(x)]\}}|N(P)\rangle
   \right],
\label{LET1}
\end{equation}
where $Q^a_5$ is the axial charge. For the $\pi^0 $ production $Q^3_5$ is involved and the commutator
with the electromagnetic current vanishes, whereas $[Q^3_5,\eta_p(x)] = -\frac12 \gamma_5 \eta_p(x)$ and
$[Q^3_5,\eta_n(x)] = \frac12 \gamma_5 \eta_n(x)$.
One obtains in this limit, e.g. for proton target,
\bea{chilimit}
 T_{\nu}^{\pi^0 p}(P,q) &\to&
\frac{i\lambda_1^p m_N}{2f_\pi}
\frac{\gamma_5(\not\!P -\not\!q + m_N)}{m_N^2-P'^2}
\left[\gamma_\nu F_1^{p}- \frac{i\sigma_{\nu\mu} q^\mu}{2m_N}F_2^{p}\right]N_p(P)\,.
\eea
%
\end{widetext}
%
Comparing this expression with the one in (\ref{rep1}) we see that the terms in $F_1^p$ and $F_2^p$ cancel out
and as the result the pion nucleon generalised form factors $G^{\pi^0 p}_1$ and $G^{\pi^0 p}_2$ are
expressed in terms of the proton magnetic and electric (Sachs) form factors, reproducing the
result in (\ref{LET}), up to corrections $\mathcal{O}(m_\pi/m_N)$.

In the sum rule language, the same result arises because the Borel-transformed correlation functions
reproduce to a good accuracy the sum rules for the $F_1^p$ and $F_2^p$ form factors in the same
approximation, i.e.
\begin{eqnarray}
 \mathbb{B}_{P'^2}[\mathcal{A}^{\pi^0 p}](M^2,Q^2)  &\simeq& \lambda_1^p e^{-m_N^2/M^2}F_1(Q^2)\,,
\nonumber\\
  \mathbb{B}_{P'^2}[\mathcal{B}^{\pi^0 p}](M^2,Q^2) &\simeq& \frac{1}{2}\lambda_1^p e^{-m_N^2/M^2}F_2(Q^2)\,,
\nonumber\\
\end{eqnarray}
so one can check that, again, the expressions in (\ref{LET}) are reproduced
up to corrections that are suppressed by powers of the pion mass. The case of $\pi^+n$ production is similar.

In Ref.~\cite{Braun:2006td} the pion production from a proton target was considered for large momentum transfers 
such that contributions of the pion-nucleon intermediate state appear to be above the continuum threshold and were dropped.
The corresponding condition is $\delta(Q^2+2m_N^2)> s_0-m_N^2$ which translates to $Q^2\ge 7.3$~GeV$^2$ for the standard value
$s_0=(1.5$~GeV$)^2$. The results are presented in \cite{Braun:2006td} in the form of
a parametrisation in terms of the axial form factor. A better way to present these results is to observe that to
the tree-level accuracy the LCSR for $G_1$ and  $G_2$ coincide with the sum rules for the
electromagnetic form factors  $F_1^p$ and  $F_2^p$, respectively, which have to be evaluated with ``chirally rotated'' nucleon
distribution amplitudes. It has to be expected, therefore, that the ratios $G_1/F_1$ and $G_1/F_2$ can be estimated more
reliably than the form factors themselves. We define, for proton target,
\begin{eqnarray}
  R_1^{\pi N} &=& Q^2 G_1^{\pi N}/(m_N^2 F_1^p)\,,
\nonumber\\
  R_2^{\pi N} &=&  G_2^{\pi N}/F_2^p\,
\end{eqnarray}
and determine $R_1^{\pi N}$ and $R_2^{\pi N}$ from the ratios of the corresponding LCSR given in \cite{Braun:2006td,Braun:2006hz}.
It turns out that the both ratios are practically constant in the relevant $Q^2 \sim 5-10$~GeV$^2$ range. Using the model for the
proton DAs suggested in Ref.~\cite{Braun:2006hz} we obtain
\begin{eqnarray}
   \hspace*{-0.5cm}R_1^{\pi^0 p} = \frac{1}{2}\,,\hspace*{1.2cm}{}&\quad&    R_2^{\pi^0 p} = -0.61(-0.64)\,,
\nonumber\\
   R_1^{\pi^+ n} = 0.88 (0.68)\,, &\quad&   R_2^{\pi^+ n} =  0.67(0.28)\,,
\label{SRratios}
\end{eqnarray}
where the numbers in parenthesis correspond to the LCSR results obtained with the asymptotic DAs. The ratio $R_1^{\pi^0 p}$ is special:
the pion-nucleon distribution amplitudes that enter the tree-level sum rule for $G_1^{\pi^0 p}$ all differ by an overall factor 1/2
from the corresponding proton DAs, apart from a numerically small off-light-cone contribution ${\mathcal O}(x^2)$, see
\cite{Braun:2006td} for the details. It follows that $R_1^{\pi^0 p} =1/2$ is a robust sum rule prediction, at tree level,
independent on the model for the nucleon DAs. The negative sign of $R_2^{\pi^0 p}$ is due to a different sign in the definition
of the $G_2^{\pi N}$ form factor in Ref.~\cite{Braun:2006td} and in Eq.~(\ref{def:amp}) as compared to the usual convention for $F_2^p$.
This ratio is also not far from 1/2, the difference being mainly the effect of the
larger off-light-cone contributions ${\mathcal O}(x^2)$ to the corresponding sum rules.

The higher sensitivity of the $\pi^+$ production form factors on the choice of the nucleon DAs should not be
considered as a drawback of the LCSR method but rather as an indication that these ratios are more sensitive to the details of the
proton structure. The main uncertainty in the given numbers is due to uncalculated
radiative corrections ${\mathcal O}(\alpha_s)$ to the LCSR.

\begin{figure}[ht]
  \includegraphics[width=0.40\textwidth,angle=0]{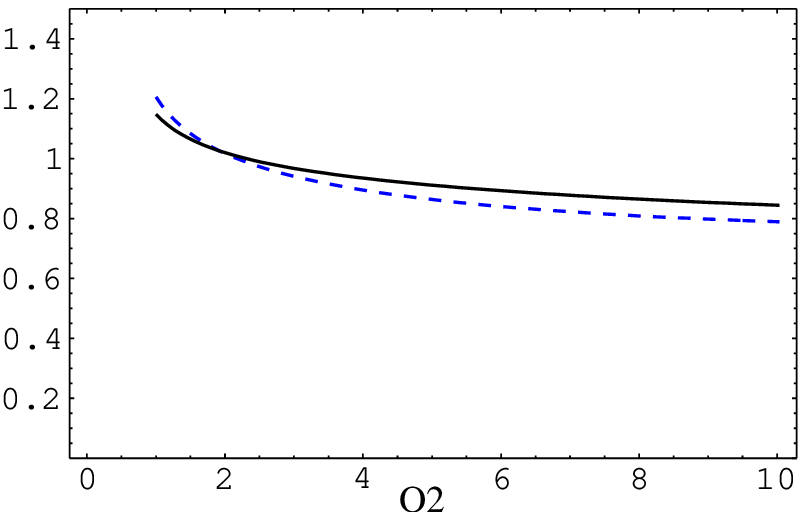}\\[2mm]
  \includegraphics[width=0.40\textwidth,angle=0]{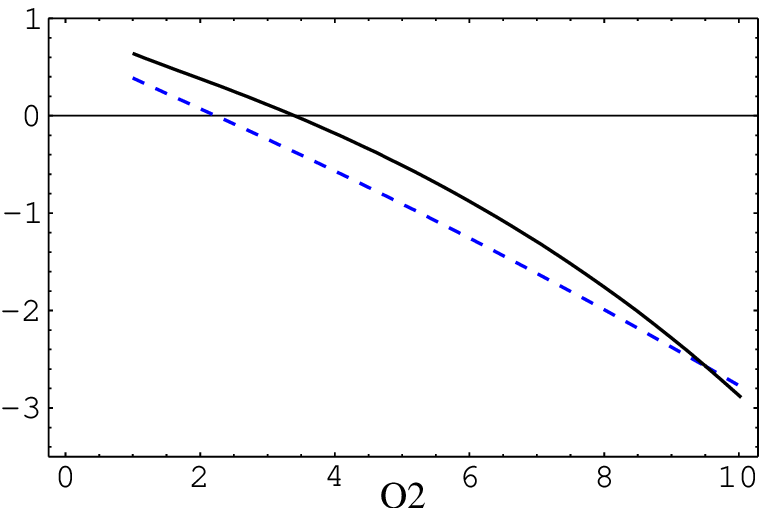}
\caption{The ratios of the LCSR predictions for the generalised form factors $G_1^{\pi^0p}$ (upper panel)
and $G_2^{\pi^0p}$ (lower panel) to the corresponding results in the soft-pion limit, Eq.~(\ref{LET}).
The solid and the dashed curves correspond to the calculation with Borel parameter $M^2=2$~GeV$^2$ and
$M^2=1$~GeV$^2$, respectively.
}
\label{fig:pi0ratio}
\end{figure}

The full sum rules in (\ref{finalSR}) essentially interpolate between the large-$Q^2$ limit considered in
\cite{Braun:2006td} and the standard prediction  based on the soft-pion theorem at low momentum transfer.
To see this, we plot in Fig.~\ref{fig:pi0ratio} the ratio of the LCSR prediction of Eq.~(\ref{finalSR}) to the
``reference model'' in Eq.~(\ref{LET}) for the form factor $G_1^{\pi^0p}$ (upper panel)
and $G_2^{\pi^0p}$ (lower panel).
{}For consistency, to make this plot we have substituted the nucleon
form factors appearing in  (\ref{finalSR}) and (\ref{LET})
by the corresponding light-cone sum rule expressions
available from Ref.~\cite{Braun:2006hz}.

The similar ratios  for $\pi^+$ production are less revealing because the corresponding LET predictions
(\ref{LET}) are very small: For $G_1^{\pi^+n}$ the contributions of the chiral rotation
and the initial state pion emission
(terms in  $G_A$ and $G_M^n$, respectively) tend to cancel each other,
whereas for $G_2^{\pi^+n}$ the initial state pion emission involves the neutron electron form factor which
is tiny. In both cases in the LCSR approach there are no superficial cancellations so that the $\pi^+n$ form factors
and generally of the same order (or bigger) than their $\pi^0p$ counterparts.


Unfortunately, at present the LCSR are only known to the leading-order accuracy in QCD perturbation theory and also the
dependence on the nucleon distribution amplitudes introduces a large uncertainty. In order to minimise this parameter
dependence we have chosen, for the purpose of this paper, to use the LCSR to determine the ratios of the Borel-transformed
correlation functions appearing in (\ref{finalSR}) to the corresponding correlation functions that enter the LCSR
for the electromagnetic form factors and take the absolute values of the form factors from experiment.
In particular we use the parametrisation of the proton magnetic form factor from \cite{Brash:2001qq} and for the
neutron magnetic form factor from \cite{Bosted:1994tm}. For the proton electric form factor we
use the fit \cite{Gayou:2001qd,Brash:2001qq} to the combined JLab data in the $0.5<Q^2<5.6$~GeV$^2$
range
\begin{eqnarray}
  \mu_p \frac{G_E^p}{G_M^p} = 1-0.13 (Q^2-0.04)
\label{GEp}
\end{eqnarray}
and put the neutron electric form factor to zero, which should be good to our accuracy.
Note that using (\ref{GEp}) for larger values of $Q^2$ up to 10 GeV$^2$ is only an extrapolation
which may be not justified.

In this way we obtain
%
\begin{widetext}
%
\bea{finalmodel}
\frac{Q^2}{m_N^2}G^{\pi^0 p}_1 &=&
 R_1^{\pi^0 p}F_1^p(Q^2)
 -\frac{1}{2}e^{-\delta(2m_N^2+Q^2)/M^2}
       \left[F_1^p(Q^2)-\frac{g_A Q^2}{Q^2+2m_N^2}G^p_M(Q^2)\right]\,,
\nonumber\\
  G^{\pi^0 p}_2 &=&
 R_2^{\pi^0 p}F_2^p(Q^2)
+e^{-\delta(2m_N^2+Q^2)/M^2}
\left[\frac{1}{2}F^p_2(Q^2) +\frac{2g_A m_N^2}{Q^2+2m_N^2}G^p_E(Q^2)\right]\,,
\nonumber\\
\frac{Q^2}{m_N^2}G^{\pi^+ n}_1 &=&
 R_1^{\pi^+ n}F_1^p(Q^2)
    -\frac{1}{\sqrt{2}}e^{-\delta(2m_N^2+Q^2)/M^2}
    \left[F^n_1(Q^2) -\frac{g_A Q^2}{Q^2+2m_N^2}G^n_M(Q^2)\right]\,,
\nonumber\\
  G^{\pi^+ n}_2 &=&
 R_2^{\pi^+ n}F_2^p(Q^2)
+ e^{-\delta(2m_N^2+Q^2)/M^2}
\left[\frac{1}{\sqrt{2}}F^n_2(Q^2) +\frac{2\sqrt{2}g_A m_N^2}{Q^2+2m_N^2}G^n_E(Q^2)\right]\,,
\eea
%
\end{widetext}
%
with the ratios $R^{\pi N}_{1,2}$ as specified in (\ref{SRratios}). 
In the rest of the calculations  we use $M^2=2$~GeV$^2$.
\begin{figure*}[ht]
  \includegraphics[width=0.80\textwidth,angle=0]{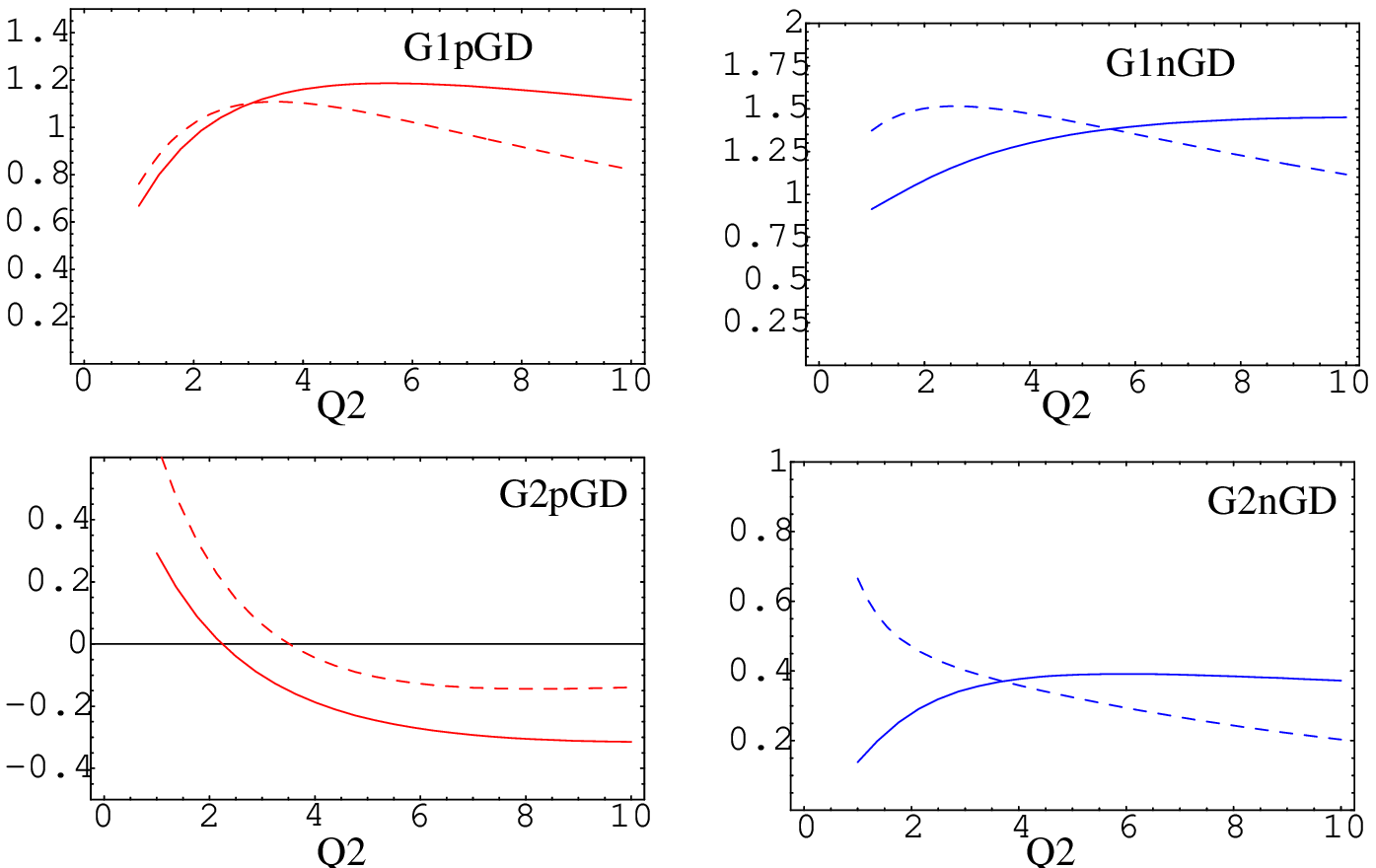}
\caption{The LCSR-based model (solid curves) for the $Q^2$ dependence of the form factors $G_{1,2}^{\pi^0 p}$
(left) and $G_{1,2}^{\pi^+ n}$ (right), (\ref{def:G12}),
normalised to the dipole formula (\ref{GD}).
The ``pure'' LCSR predictions (all form factors and other input taken directly from the sum rules) are 
shown by the dashed curves for comparison.
}
\label{fig:G12proton}
\end{figure*}
\begin{figure*}[ht]
  \includegraphics[width=0.80\textwidth,angle=0]{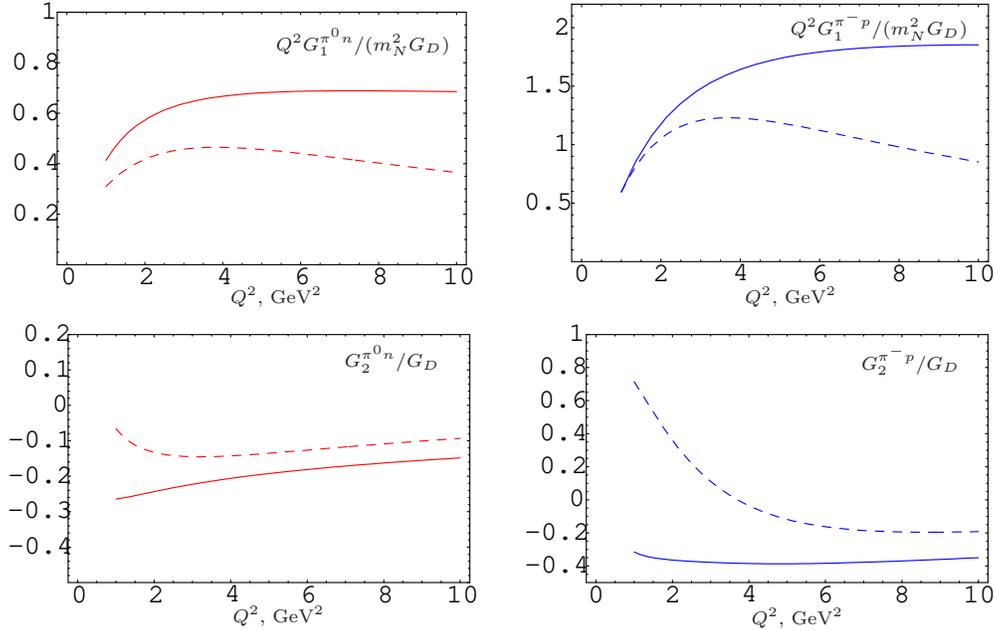}
\caption{The LCSR-based model (solid curves) for the $Q^2$ dependence of the form factors $G_{1,2}^{\pi^0 n}$
(left) and $G_{1,2}^{\pi^- p}$ (right), normalised to the dipole formula (\ref{GD}).
The ``pure'' LCSR predictions (all form factors and other input taken directly from the sum rules) are 
shown by the dashed curves for comparison.
}
\label{fig:G12neutron}
\end{figure*}

The results are shown by solid curves in Fig.~\ref{fig:G12proton}. For comparison, ``pure'' LCSR predictions
(all form factors and other input taken directly from the sum rules) are shown by the dashed curves. 
In general, $G_1$ form factors can be predicted more reliably than  
$G_2$ as the latter are more sensitive to higher twist corrections to the sum rules.  
The large difference between solid and dashed curves for $G_{2}^{\pi^+ n}$, at small $Q^2$ is due to strong 
cancellations among various contributions.

The LCSR for the pion electroproduction from the neutron target
$e(l)+n(P) \to e(l') + \pi^0(k) + n(P')$\,,
$e(l)+n(P) \to e(l') + \pi^-(k) + p(P')$,
can be obtained from the expressions given in \cite{Braun:2006td} by the substitution $e_u \leftrightarrow e_d$.
Following the same procedure as for the proton, we define the ratios of $G_{1,2}^{\pi^0 n}$ and $G_{1,2}^{\pi^- p}$ to
the neutron Dirac and Pauli form factors at large $Q^2$ where contributions of the pion-nucleon intermediate state can be omitted as
\begin{eqnarray}
  R_1^{\pi N} &=& Q^2 G_1^{\pi N}/(m_N^2 F_1^n)\,,
\nonumber\\
  R_2^{\pi N} &=&  G_2^{\pi N}/F_2^n\,
\end{eqnarray}
and determine $R_{1,2}^{\pi^0 n}$ and $R_{1,2}^{\pi^- p}$ from the ratios of the corresponding LCSR.
Using the model for the nucleon DAs suggested in Ref.~\cite{Braun:2006hz} we obtain
\begin{eqnarray}
   \hspace*{-0.5cm}R_1^{\pi^0 n} = -\frac{1}{2}\,,\hspace*{1.2cm}{}&\quad&    R_2^{\pi^0 n} = 0.58(0.57)\,,
\nonumber\\
   R_1^{\pi^- p} = -1.37 (-0.74)\,, &\quad&   R_2^{\pi^- p} = 1.32(0.32)\,,
\nonumber\\
\label{neutron2}
\end{eqnarray}
which is the counterpart  of Eq.~(\ref{SRratios}). 
Note that the normalisation is in the present case to the neutron electromagnetic form factors, not the proton ones.
The numbers in parenthesis correspond to the LCSR results obtained with the asymptotic DAs.
The complete LCSR-based model is then constructed using these ratios and adding the contributions of pion-nucleon states,
in full analogy with Eq.~(\ref{finalmodel}), 
with obvious substitutions proton$\leftrightarrow$neutron in the form factors that are involved:
%
\begin{widetext}
%
\bea{finalmodelneutron}
\frac{Q^2}{m_N^2}G^{\pi^0 n}_1 &=&
 R_1^{\pi^0 n}F_1^n(Q^2)
 +\frac{1}{2}e^{-\delta(2m_N^2+Q^2)/M^2}
       \left[F_1^n(Q^2)-\frac{g_A Q^2}{Q^2+2m_N^2}G^n_M(Q^2)\right]\,,
\nonumber\\
  G^{\pi^0 n}_2 &=&
 R_2^{\pi^0 n}F_2^n(Q^2)
-e^{-\delta(2m_N^2+Q^2)/M^2}
\left[\frac{1}{2}F^n_2(Q^2) +\frac{2g_A m_N^2}{Q^2+2m_N^2}G^n_E(Q^2)\right]\,,
\nonumber\\
\frac{Q^2}{m_N^2}G^{\pi^- p}_1 &=&
 R_1^{\pi^- p}F_1^n(Q^2)
    -\frac{1}{\sqrt{2}}e^{-\delta(2m_N^2+Q^2)/M^2}
    \left[F^p_1(Q^2) -\frac{g_A Q^2}{Q^2+2m_N^2}G^p_M(Q^2)\right]\,,
\nonumber\\
  G^{\pi^- p}_2 &=&
 R_2^{\pi^- p}F_2^n(Q^2)
+ e^{-\delta(2m_N^2+Q^2)/M^2}
\left[\frac{1}{\sqrt{2}}F^p_2(Q^2) +\frac{2\sqrt{2}g_A m_N^2}{Q^2+2m_N^2}G^p_E(Q^2)\right]\,.
\eea
%
\end{widetext}
%
The results are shown in Fig.~\ref{fig:G12neutron} (solid curves). The ``pure'' LCSR predictions
(all form factors and other input taken directly from the sum rules) are shown by the dashed curves
for comparison.

\end{document}